\PassOptionsToPackage{prologue,table}{xcolor}
\documentclass[acmtog,screen]{acmart}

\acmSubmissionID{1138}

\PassOptionsToPackage{prologue,table,xcdraw}{xcolor}
\usepackage{xargs}
\usepackage{booktabs} %
\usepackage{tabularx}
\usepackage{caption}
\usepackage{nicematrix}
\usepackage{float}
\usepackage{mathtools}
\usepackage{amsmath,amsfonts}
\usepackage{bm}
\usepackage{microtype}
\usepackage{units}
\usepackage{subcaption}
\usepackage{overpic}
\usepackage{cancel}
\usepackage{wrapfig}
\usepackage{placeins}
\usepackage{empheq}
\usepackage[normalem]{ulem}
\usepackage[export]{adjustbox}
\usepackage[vlined,boxed,ruled]{algorithm2e} %
\usepackage{tikz}
\usepackage{xstring}
\usepackage{xr}
\usepackage{multirow}
\usepackage[capitalize,noabbrev]{cleveref} %
\usepackage{url} %
\usepackage[utf8]{inputenc}
\usepackage{enumitem}
\usepackage{wrapfig}
\usepackage{listings}
\usepackage[frozencache]{minted}
\usepackage{xfrac}
\usepackage{relsize}
\usepackage{fvextra} %
\usepackage{adjustbox}

\setlist[itemize]{noitemsep, nolistsep, leftmargin=*}
\setlist[enumerate]{noitemsep, nolistsep, leftmargin=*}

\SetKw{Or}{or}
\SetKw{And}{and}
\SetStartEndCondition{ }{}{}%
\SetKwProg{Fn}{def}{\string:}{}
\SetKwFunction{Range}{range}%
\SetKw{KwTo}{in}
\SetKwFor{For}{for}{\string:}{}%
\SetKwFor{While}{while}{:}{fintq}%
\SetKwIF{If}{ElseIf}{Else}{if}{:}{elif}{else:}{}%
\SetKwComment{Comment}{\# }{}
\SetKw{Continue}{continue}
\AlgoDontDisplayBlockMarkers
\SetAlgoNoEnd
\SetAlgoNoLine
\DontPrintSemicolon

\SetAlFnt{\small}
\SetAlCapFnt{\small}
\SetAlCapNameFnt{\small}
\SetAlCapHSkip{0pt}
\IncMargin{-\parindent}

\definecolor{maroon}{cmyk}{0, 0.87, 0.68, 0.32}
\definecolor{halfgray}{gray}{0.55}
\definecolor{ipython_frame}{RGB}{207, 207, 207}
\definecolor{ipython_bg}{RGB}{247, 247, 247}
\definecolor{ipython_red}{RGB}{186, 33, 33}
\definecolor{ipython_green}{RGB}{0, 128, 0}
\definecolor{ipython_cyan}{RGB}{64, 128, 128}
\definecolor{ipython_purple}{RGB}{170, 34, 255}

\newcommand{\ccomment}[1]{\textcolor{ipython_cyan}{#1}}
\setminted{
  bgcolor=ipython_bg,
  rulecolor=\color{ipython_frame},
  fontsize=\scriptsize,
  linenos,
  numbersep=2mm,
  breaklines,
  breakanywhere,
  tabsize=1,
  encoding=utf8,
  escapeinside=||,
}
\usetikzlibrary{shadows} %
\usepackage[framemethod=tikz]{mdframed} %

\newmdenv[
    linecolor=ipython_frame,
    linewidth=0.4pt,
    roundcorner=5pt,
    backgroundcolor=ipython_bg,
    shadow=true,
    shadowsize=3pt,
    shadowcolor=black!30,
    innerleftmargin=1mm,
    innerrightmargin=1mm,
    innertopmargin=0mm,
    innerbottommargin=0mm,
    skipabove=0.5\topskip,
    skipbelow=0.5\topskip,
    nobreak=true
]{roundedminted}

\usepackage{etoolbox}
\BeforeBeginEnvironment{minted}{\begin{roundedminted}}
\AfterEndEnvironment{minted}{\end{roundedminted}}

\definecolor{ListBGColor}{rgb}{0.95,0.95,0.95}
\definecolor{ListCommentColor}{rgb}{0.3,0.5,0.3}
\definecolor{ListKeywordColor}{rgb}{0.2,0.2,0.7}
\definecolor{ListStringColor}{rgb}{0.6,0.1,0.1}
\definecolor{ListNumberColor}{rgb}{0.4,0.4,0.4}
\definecolor{ListIdentifierColor}{rgb}{0.1,0.1,0.1}
\def\commentType{0}

\ifnum\commentType=0
    \newcommandx{\customComment}[3]{}
    \newcommandx{\customTODO}[3]{}
\fi
\ifnum\commentType=1
    \usepackage[
    type=upperleft,
    unit=in
    ]{fgruler}
    \newcommandx{\customComment}[3]{\textbf{\scriptsize \color{#2} #1: #3}}
    \newcommandx{\customTODO}[3]{\textbf{\scriptsize \color{#2} #1: #3}}
\fi
\ifnum\commentType=2
    \usepackage[
    type=upperleft,
    unit=in
    ]{fgruler}
    \usepackage{pdfcomment}
    \newcommandx{\customComment}[3]{\pdfcomment[icon=Comment,opacity=0.5,color=#2,author=#1]{#3}}
    \newcommandx{\customTODO}[3]{\pdfcomment[icon=Note,opacity=0.5,color=#2,author=#1]{#3}}
\fi
\ifnum\commentType=3
    \usepackage[
    type=upperleft,
    unit=in
    ]{fgruler}
    \usepackage{pdfcomment} %
    \usepackage{todonotes} %
    \usepackage{silence}
    \newcommandx{\customComment}[3]{\todo[color=#2!40,size=\small]{\textbf{#1:} #3}}
    \newcommandx{\customTODO}[3]{\todo[color=#2!40,size=\small]{\textbf{#1:} #3}}
\fi

\let\originalleft\left %
\let\originalright\right %
\renewcommand{\left}{\mathopen{}\mathclose\bgroup\originalleft} %
\renewcommand{\right}{\aftergroup\egroup\originalright} %

\definecolor{amber}{rgb}{1.0, 0.49, 0.0}
\definecolor{darkgreen}{rgb}{0.0, 0.5, 0.0}
\definecolor{darkblue}{rgb}{0.0, 0.0, 0.5}
\definecolor{darkred}{rgb}{0.5, 0.0, 0.0}
\definecolor{bloodred}{rgb}{0.616, 0.133, 0.208}
\definecolor{forestgreen}{rgb}{0.133, 0.616, 0.541}
\definecolor{babyblue}{RGB}{91, 206, 250}
\definecolor{babypink}{RGB}{245, 169, 184}
\definecolor{ahmedcolor}{RGB}{255, 92, 255}

\newcommandx{\All}[1]{\customComment{All}{red}{#1}}
\newcommandx{\ana}[1]{\customComment{AD}{babyblue}{#1}}
\newcommandx{\vincent}[1]{\customComment{VS}{forestgreen}{#1}}
\newcommandx{\silvia}[1]{\customComment{Silvia}{babypink}{#1}}

\newcommandx{\TODO}[1]{\customTODO{TODO}{red}{#1}}
\newcommandx{\todo}[1]{\customTODO{TODO}{red}{#1}}

\newcommand{\REMOVE}[1]{} %

\def\equationautorefname~#1\null{%
  Equation~(#1)\null
}

\newcommand{\suchThat}[0]{{\mathrm{s.t.}}}

\newcommand{\tpose}[0]{\top}  %
\renewcommand{\vec}[1]{\boldsymbol{#1}}  %
\let\origEll\ell
\renewcommand{\ell}[0]{\mathlarger{\origEll}}
\let\oldcdot\cdot
\renewcommand{\cdot}[0]{{\;\oldcdot\;}}
\newcommand{\R}[0]{{\mathbb{R}}}

\newcommand{\N}{\mathbb{N}}

\newcommand{\FS}{\mathbb{F}}
\newcommand{\TS}{\mathbb{T}}
\newcommand{\CS}{\mathbb{C}}
\newcommand{\ES}{\mathbb{E}}
\newcommand{\VS}{\mathbb{V}}

\newcommand{\HS}{\mathbb{H}}

\newcommand{\CT}{\textsc{CT}}

\newcommand{\EV}{\textsc{EV}}
\newcommand{\VV}{\textsc{VV}}

\newcommand{\Cv}{{\textsc{C}_V}}
\newcommand{\Tv}{{\textsc{T}_V}}
\newcommand{\Ev}{{\textsc{E}_V}}

\DeclareMathAlphabet{\mathmybb}{U}{bbold}{m}{n}

\newcommand{\Diff}[1]{\mathrm{d}#1}

\newcommand{\diag}{\mathop{\mathrm{diag}}}

\DeclareMathOperator{\Tr}{Tr}

\DeclareMathOperator{\vectorize}{vec}

\DeclareMathOperator*{\argmin}{arg\,min}

\DeclareMathOperator*{\relu}{relu}

\definecolor{cartoPrismTeal}{rgb}{0.21960784 0.65098039 0.64705882}
\definecolor{cartoPrismOrange}{rgb}{0.88235294 0.48627451 0.01960784}
\definecolor{cartoPrismGreen}{rgb}{0.45098039 0.68627451 0.28235294}
\definecolor{cartoPrismRed}{rgb}{0.8 0.31372549 0.24313725}
\definecolor{cartoPrismPurple}{rgb}{0.58039216 0.20392157 0.43137255}

\definecolor{mathematicaBlue}{rgb}{0.38, 0.51, 0.71}
\definecolor{mathematicaOrange}{rgb}{0.88, 0.61, 0.14}
\definecolor{mathematicaGreen}{rgb}{0.56, 0.69, 0.19}
\definecolor{mathematicaRed}{rgb}{0.92,0.39, 0.21}
\definecolor{mathematicaPurple}{rgb}{0.53, 0.47, 0.7}

\acmJournal{TOG}
\acmArticle{112}
\setcopyright{cc}
\setcctype{by-nc-nd}
\acmYear{2026}
\acmVolume{45}
\acmNumber{4}
\acmMonth{7}
\acmDOI{10.1145/3811371}

\begin{document}

\title[\texttt{iskra}: A System for Inverse Geometry Processing]{\texttt{iskra}: A System for Inverse Geometry Processing}

\author{Ana Dodik}
\email{anadodik@mit.edu}
\orcid{0000-0003-4391-8877}
\affiliation{%
  \institution{MIT CSAIL}
   \streetaddress{32 Vassar St}
  \city{Cambridge}
  \state{MA}
  \postcode{02139}
  \country{USA}  
}

\author{Ahmed H.\ Mahmoud}
\email{ahdhn@mit.edu}
\orcid{0000-0003-1857-913X}
\affiliation{%
  \institution{MIT CSAIL}
  \streetaddress{32 Vassar St}
  \city{Cambridge}
  \state{MA}
  \postcode{02139}
  \country{USA}
}

\author{Justin Solomon}
\email{jsolomon@mit.edu}
\orcid{0000-0002-7701-7586}
\affiliation{%
  \institution{MIT CSAIL}
  \streetaddress{32 Vassar St}
  \city{Cambridge}
  \state{MA}
  \postcode{02139}
  \country{USA}
}


\begin{abstract}

We propose a system for differentiating through solutions to geometry processing problems.
Our system differentiates a broad class of geometric algorithms, exploiting existing fast problem-specific schemes common to geometry processing, including local-global and ADMM solvers.
It is compatible with machine learning frameworks, opening doors to new classes of inverse geometry processing applications.
We marry the scatter-gather approach to mesh processing with tensor-based workflows and rely on the adjoint method applied to user-specified imperative code to generate an efficient backward pass behind the scenes.
We demonstrate our approach by differentiating through mean curvature flow, spectral conformal parameterization, geodesic distance computation, and as-rigid-as-possible deformation, examining usability and performance on these applications.
Our system allows practitioners to differentiate through existing geometry processing algorithms without needing to reformulate them, resulting in low implementation effort, fast runtimes, and lower memory requirements than differentiable optimization tools not tailored to geometry processing.

\end{abstract}

\begin{teaserfigure}
  \centering
  \includegraphics[width=\linewidth]{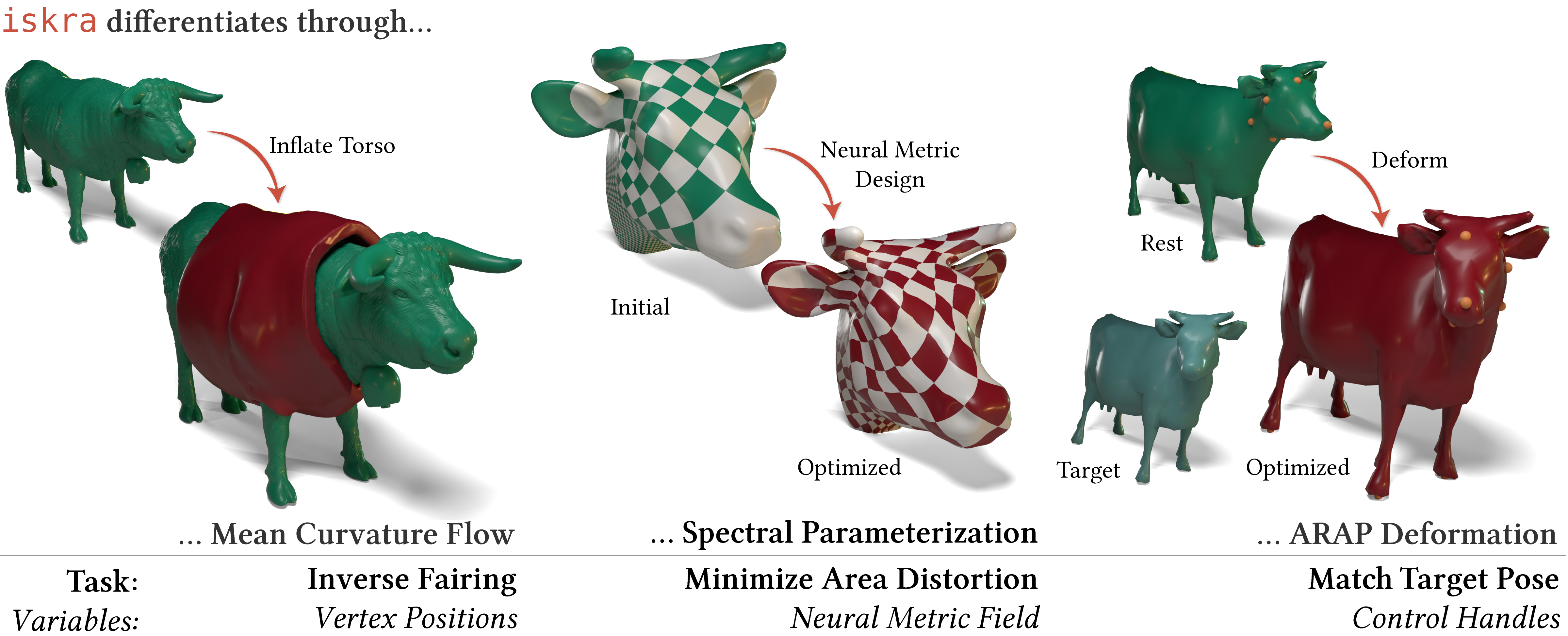}
  \caption{Our system embeds geometry processing algorithms within larger differentiable pipelines. Users can optimize vertex positions to run a fairing flow backwards, a mesh's intrinsic metric to reduce area distortion during parameterization, or control handles to match a deformed mesh to a target pose.}
  \Description{}\label{fig:teaser}
\end{teaserfigure}

\maketitle

\section{Introduction}\label{sec:intro}

Much of geometry processing consists of optimization problems over mesh data that are easily expressed in mesh-specific differentiable systems~\cite{herholz2022sparsity, herholz2024mesh,mara2021thallo}.
For example, these systems address the classical problem of deforming a mesh by minimizing a distortion measure while constraining a set of control vertices.
However, what if we wish to use the deformed solution as part of a larger program, e.g., finding the positions of the controls such that the deformed mesh best approximates some target?
We call this an \emph{inverse problem}, consisting of an \emph{inner optimization} that solves for the deformed mesh and an \emph{outer optimization} which requires the derivatives of deformed mesh with respect to the control vertices (Figure~\ref{fig:arap}).

Inverse geometry processing problems introduce two coupled challenges that make existing systems unsuitable for this case.
The first involves operating on mesh data, which requires system support for sparse and irregular operations and workloads.
The second has to do with the diversity of optimization problems and problem-specific solvers found in geometry processing. 
Besides standard sparse linear systems, geometry processing software often solves highly non-linear or non-convex problems using specialized methods that leverage structures specific to the problem.

Given these requirements, existing tools for inverse problems imply trade-offs.
The first class of tools offers simple \emph{declarative} interfaces for specifying the forward problem in some canonical form, which they automatically optimize using a pre-established solver.
These tools leverage the narrow canonical form of the inner problem to provide derivatives for the outer problem~\cite{agrawal2019cvxpylayers,pineda2022theseus,lai2023deltaprox}. 
This paradigm offers a concise interface for the class of problems these system target at the cost of constraining the problem and solver class.
Going back to our initial example, solving the seminal as-rigid-as-possible (ARAP) deformation problem~\cite{Sorkine:ARAP, Igarashi:ARAP} with quadratic programming systems~\cite{Pan:2024:BPQP} or generic optimization libraries~\cite{pineda2022theseus} foregoes the popular alternating algorithm typically used to optimize ARAP problems~\citet{Sorkine:ARAP} and may not be possible unless the (nonconvex) ARAP objective can be massaged into an acceptable canonical form.  %
Thus, this class of systems prevents the user from relying on specialized solvers that exploit the problem's structure.

Alternatively, \emph{imperative tools} offer a lower-level interface, allowing users to specify optimization loops themselves.
Here, one might be tempted to simply rely on an off-the-shelf framework, e.g., PyTorch or Jax, to implement an inner optimization loop, which the framework will ``unroll'' during the backward pass to provide derivatives for the outer problem.
Alas, this process stores intermediate data with each iteration, quickly blowing up the memory for many problems.
Moreover, it prevents us from relying on highly-optimized black-box solvers for the forward pass that are not implemented in the framework, e.g., ARPACK for eigenproblems~\cite{Lehoucq:1998:AUG}.
Be they declarative or imperative, support for sparse computation on meshes is rare in systems for inverse problems, as many are meant for domains like machine learning or image processing where sparsity is not a key consideration~\cite{Li:2018:DPI, lai2023deltaprox,chen2018neuralode}.

In bringing support for sparse computation on meshes to inverse problem systems, we must choose a programming abstraction---see Section~\ref{sec:related} for an overview of existing options.
Representing mesh data in a dense memory layout and exposing it to the user via the tensor abstraction would come with major benefits: it would enable rapid prototyping and make the boundaries to numerical and machine learning ecosystems porous, e.g.\ via the \texttt{DLPack} interface.
Such interoperability would be crucial for inverse geometry processing specifically, as it would allow the outer optimization loop to be anything ranging from inverse rendering of the mesh, to neural network dictated mesh variables or optimization objectives.
At the same time, relying on this abstraction is not entirely straight-forward and would necessitate designing high-level user abstractions for sparse computation patterns common to meshes (\S\ref{sec:geo_module}).

Taking stock of constraints, we distill the following design goals:
\begin{itemize}
    \item \textbf{Versatility.} Our system must support differentiating through the broad spectrum of geometry processing tasks and offer flexibility in how they are solved and differentiated.
    \item \textbf{Usability.} It must integrate into a clean interface for differentiable programming, geometry processing, and inverse problems.
    \item \textbf{Interoperability.} Geometric problems must compose seamlessly with  numerical computing and machine learning ecosystems.
    \item \textbf{Efficiency.} Our system should be sparsity-aware and integrate with both CPU and GPU workflows.
\end{itemize}

To this end, we introduce \texttt{iskra}, an \emph{imperative} tensor-based system for inverse geometry processing.
It offers a high-level interface for differentiable geometry processing, staying interoperable with the broader numerical computing ecosystem and offering sparse CPU or GPU execution.
Our system introduces a simple high-level interface that allows the user to differentiate through common geometry processing solvers.
Practically, this means that \texttt{iskra} allows geometric problems to be embedded in machine learning pipelines (\S\ref{ss:scp}). %
Our system supports differentiating through common geometry processing tasks and solvers including sparse linear systems, local-global solvers, sparse eigenproblems (\S\ref{sec:implicit_module}), as well as more esoteric methods that exploit problem-specific structures (\S\ref{ss:arap}, \S\ref{ss:geodesics}).

In particular, there is a rich body of work on specialized solvers \cite{manson2011hierarchical, bourquat2022hierarchical, Solomon:2015:CWS, Solomon:2014:EMD}, reflecting the collective knowledge and expertise of the past $30$ years of the geometry processing literature. 
The imperative nature of our system allows users to differentiate through geometry problems without reformulating seminal approaches.
The user must only specify one iteration of a fixed-point or root-finding problem, which our system then differentiate through using the \emph{adjoint method} (\S\ref{sec:implicit_module}).
Thus, to avoid locking the user into one specific approach, we offer flexibility in separating forward and backward pass implementation as well as the specifics of each.

We provide a clean interface that integrates geometric primitives with the tensor-based abstraction.
Since it is built on top of PyTorch, our system has first-class interoperability with machine learning ecosystems and follows standard practices that enable CPU/GPU integration.
To make the sparse geometry workflows compatible with a dense tensor framework, our system exposes a set of mesh-specific \emph{gather-scatter} primitives and offers custom data types and operators for differentiable sparse linear algebra.
It also offers an extensive list of routines needed to build geometry processing algorithms (\S\ref{sec:geo_module}).

We demonstrate the versatility of our system on a set of problems, each of which requires differentiating through a different kind of solver.
These include a geometric flow~\cite{desbrun2023fairing} with a sparse linear system (\S\ref{ss:mcf}), mesh parameterization~\cite{Mullen:2008:SCP} with a sparse eigensystem (\S\ref{ss:scp}), mesh deformation with a highly nonlinear and non-convex problem~\cite{Sorkine:ARAP} solved via a custom local-global approach (\S\ref{ss:arap}), and geodesic computation~\cite{edelstein2023rgd} using a custom ADMM-based~\cite{boyd2011distributed} optimizer (\S\ref{ss:geodesics}).
Our implementation is open-source and available at \url{https://github.com/anadodik/iskra}.

In this paper, we make the following contributions:
\begin{itemize}
    \item a system for efficient inverse geometry processing that integrates cleanly with differentiable and numerical computing ecosystems;
    \item a robust suite of differentiable geometry operators and optimizers with flexible way of specifying the backward pass;
    \item a set of diverse example applications demonstrating practicality, performance, and ease of use; and
    \item a quantitative evaluation and benchmark of the system, its various components, and user trade-offs.
\end{itemize}

\paragraph*{Scope} %
Our system design reflects certain trade-offs.
Our goal is to balance efficiency while prioritizing interoperability, generality and usability, not to aim for peak performance or memory utilization.
Moreover, we limit our scope to exclude dynamic changes in sparsity or mesh connectivity during solver iterations, as well as physics simulation-specific considerations such as collision detection or contact handling.
We leave for future work to integrate with \texttt{iskra} low-level approaches to geometry, dynamic topology changes, and physics-specific considerations

\section{Related Work}\label{sec:related}

Prior work spans two largely orthogonal research directions. The first is systems for differentiable geometry processing aiming to integrate automatic differentiation into geometry processing with focus on performance, flexibility, and interoperability. The second focuses on differentiating through optimization problems more broadly by developing general frameworks for backpropagation through solvers. We review these lines of work separately as they address complementary aspects of the challenges our system solves. We also review works on differentiating through ODE and PDE solvers. 

\subsection{Geometry Processing Differentiation Systems}\label{ss:diffgp}
Contemporary geometry processing software for triangle and tetrahedral meshes falls into one of two categories. The first, exemplified by the seminal \texttt{libigl} library, offers high-level interfaces based on the matrix abstraction, focusing on practicality and ease of use and offering a rich geometry processing toolbox~\cite{cgal:eb-25b, libigl, gptoolbox, gpytoolbox, Botsch:2002:OAG, directional}. This set of tools often leaves out differentiability and GPU acceleration necessary for machine learning and inverse problem interoperability.

The second camp focuses on speed, often via GPU acceleration. These GPU-based systems rely on low-level abstractions that require the user to specify every operation over a local neighborhood of a mesh element (e.g., for each vertex, perform an operation on its neighboring triangles)~\cite{yu2023taichi, mahmoud2021rxmesh, herholz2024mesh, mahmoud2025dynamic}. Such low-level abstractions lead to inflexibility and incompatibility with tensor abstractions used within the machine learning ecosystem~\cite{mahmoud2021rxmesh, yu2023taichi}.

Recently, several attempts have been proposed to introduce differentiability as a first-class primitive for triangle and tetrahedral mesh processing. TinyAD~\cite{schmidt2022tinyad} made the observation that forward-mode automatic differentiation is faster for geometry processing workload. Such a system can be easily integrated with \texttt{libigl} and OpenMesh~\cite{Botsch:2002:OAG} to make them differentiable. However, TinyAD lacks GPU acceleration and does not integrate with modern machine learning frameworks.  Other systems with similar limitations include \citet{herholz2022sparsity, herholz2024mesh, mara2021thallo, Fernandez:2025:SEB}, which offer efficient differentiation but provide limited support for interoperability with ML ecosystems. Crucially, since the goal of these systems is to differentiate in service of solving geometry processing problems, they only allow for differentiability over a individual operations over mesh-based data, which is insufficient if we wish to \emph{invert} a geometry processing problem, i.e., differentiate through solutions to geometric optimizations (see \S\ref{sec:background}).

More broadly, existing solutions for differentiable geometry processing fall into two categories: (1)~general-purpose automatic differentiation systems that are flexible but inefficient for sparse geometric operators (e.g., PyTorch, JAX) and (2)~efficient but narrowly-scoped systems that lack GPU acceleration, ML interoperability, or support for differentiable optimization. In contrast to both extremes, our work focuses on usability, interoperability, and abstraction flexibility, rather than focusing solely on raw differentiation speed or mesh processing throughput.

While systematic approaches to \emph{differentiable} geometry processing are well researched, to our knowledge there are no systems specifically focusing on \emph{inverse} geometry processing, i.e., the systematic differentiation through geometric optimizations.
Nonetheless, the geometry processing literature has already made use of differentiable optimizations in an ad-hoc manner.
Existing work in geometry that differentiates through linear solvers can be found in \citet{wang2021fast, wang2023maps}, through eigenproblems in \citet{cosmo2019isospec,smirnov2021hodgenet}, and through geodesic distances in  \citet{verninas2026disgeod, li2024geodesics}.%

\subsection{Systems for Differentiation Through Optimization}
Many systems allow differentiation through optimization routines, i.e., backpropagating through solutions of optimization problems. For example, Theseus~\cite{pineda2022theseus} is an application-agnostic framework for differentiable nonlinear least squares. Like our system, it strives for interoperability within the machine learning ecosystem. However, because it is not specialized for geometry processing, its performance degrades when differentiating through complex operations such as sparse linear solvers or eigensolvers.

\textsc{CVXPYLayers}~\cite{agrawal2019cvxpylayers} is a domain-specific language that differentiates through disciplined convex programs by converting the problem specification into certain canonical forms. From there, \textsc{CVXPYLayers} differentiates through these problems without explicitly backpropagating through the operations of the solver. Instead, they use implicit differentiation formulas specific to canonicalized versions of certain convex problems. %
While this facilitates backpropagation, the user must define problems as disciplined convex programs, which might not be always possible. Even when possible, their system does not scale, even to moderate-sized meshes (\S\ref{ss:geodesics}).

Similarly to \textsc{CVXPYLayers}, $\nabla$-Prox~\cite{lai2023deltaprox} introduces a domain-specific language and compiler for differentiable proximal algorithms, targeted mainly for large-scale image optimization. $\nabla$-Prox reduces a high-level problem specification to a canonical proximal form and compiles it into a differentiable solver that supports unrolling, implicit differentiation, and hybrid learning schedules. Their design enables efficient differentiation through iterative proximal algorithms and allows for rapid prototyping. However, the system is tailored to dense, regular image operations and does not address sparse, irregular operations arising in geometry processing, specifically on triangle and tetrahedral meshes.

BPQP~\cite{Pan:2024:BPQP} is a differentiable convex optimization layer that reformulates the backward pass through a convex program as a simplified quadratic program by exploiting the structure of the KKT conditions. BPQP computes gradients via a first-order QP\@. This reformulation reduces the cost of implicit differentiation compared to unrolling the forward solver or differentiating through a full KKT system. Prior to BPQP, OptNet~\cite{amos2017optnet} differentiates through quadratic programs by implicitly differentiating the full KKT system associated with a primal--dual interior-point solve, making OptNet's backward pass relatively straightforward.

\subsection{Differentiation Through ODEs and PDEs}

A complementary line of work to ours enables differentiation through continuous time/space of dynamical systems. These systems  backpropagate through the solution of ordinary and partial differential equations, implementing the simulator as a layer that can be differentiated implicitly. 
For example, Neural ODE~\cite{chen2018neuralode} backpropagates through black-box ODE solvers using the adjoint method, avoiding backpropagation through steps of the solver to enable end-to-end training of models defined by continuous dynamics.
For more information, refer to \citet{kolter2020deepimplicit}.

Subsequent systems such as ChainQueen~\cite{Hu:2019:CQA} and  DiffTaichi~\cite{hu2019difftaichi} extended this idea to differentiable physics simulation by relying on compilers to support backpropagating through PDE-based solvers for fluids, soft bodies, and rigid-body dynamics---notably with no support for meshes.
Systems like DiffPD \cite{du2021diffpd} and that of \citet{huang2024diff} provide differentiable systems specially tailored to soft-body simulation.
Much like the differentiable optimization systems, these methods offer a fixed set of pre-chosen soft-body solvers \cite{du2021diffpd, huang2024diff}.
For information about recent advances on differentiable physics simulators, see \citet{Coros:2021:DS,Newbury:2024:ARO}. 

These systems enable differentiating through time-dependent soft-body solvers on \emph{volumetric} domains with collision handling, and primarily target optimizing for physical parameters.
In contrast, our focus is on inverse geometry processing over static triangle and tetrahedral meshes.
Geometry processing problems do not fit neatly into a unified declarative form like elasticity does \cite{Hu:2019:CQA, du2021diffpd, huang2024diff}.
Even if these systems were modified to support codimension-$1$ meshes, expressing surface ARAP as a general nonlinear optimization problem requires reformulating the objective function, much like for Theseus~\cite{pineda2022theseus} (\S\ref{ss:arap}).
Moreover, two of our examples---eigenproblems (\S\ref{ss:scp}) and geodesics on curved domains (\S\ref{ss:geodesics})---are not handled by prior systems.
Lastly, it is common that the systems in this class do not permit specifying custom forward solvers~\cite{Hu:2019:CQA, du2021diffpd, huang2024diff}; in contrast, ours tackles general inverse problems, including linear systems, eigenproblems, fixed points, and proximal iterations.
As a result, the abstractions and optimizations developed for differentiable simulation do not directly address the sparsity in geometry processing, nor the broader class of geometry processing problems that our system supports.

\section{Background}\label{sec:background}
Inverse geometry processing requires differentiating through a broad variety of optimization problems.
This section briefly summarizes the mathematical backbone of our system, namely implicit differentiation and its implementation via the adjoint method.
For more information on the topic, see the textbook by \citet{strang2007computational}.

\subsection{Adjoint Method}\label{sec:implicit_diff}

Suppose $g(\cdot)$ denotes a solver for a geometry processing problem with parameters $x$; we will use $x\in\R^m$ to denote the input to $g(\cdot)$ and $y\in\R^n$ to denote the output $y=g(x)$. 
As an example, we may be interested in finding the sensitivites of deformed vertex coordinates ($y$) to changes in control handles ($x$) in as-rigid-as-possible deformation.
The adjoint method allows us to calculate these derivatives under mild conditions, given that we can write down a differentiable function ${f : \R^m \times \R^n \to \R^n}$ implicitly relating inputs $x$ of $g$ to outputs $y$:
\begin{equation}~\label{eq:implicit-fun}
    f(x; y) = 0,
\end{equation}
where we assume that root-finding on $f$ \emph{uniquely} determines $y$ from $x$.
We use this notation without loss of generality, assuming matrix and tensor valued $x$ and $y$ are vectorized.
For notational convenience, we allow for multiple variables on either side of the semicolon in $f$, implicitly assuming they are concatenated to form the two vectors $x$ and $y$; we always differentiate the symbols after the semicolon with respect to those before it.

Using gradient-based optimization of objectives that involve $g$ requires finding the Jacobian ${\sfrac{\partial y}{\partial x} \in \R^{n \times m}}$.
Rather than working directly with $g$, which is typically a complicated map, 
the implicit function theorem states that if the Jacobian $\sfrac{\partial f}{\partial y} \in \R^{n \times n}$ is invertible at an $x$, then $y$ can be locally treated as a function of $x$, denoted $y(x)$.
By the chain rule, taking a total derivative of $f(x; y(x)) = 0$ with respect to $x$ yields
\begin{equation*}
    \frac{\Diff f}{\Diff x}
    =
    \frac{\partial f}{\partial x}
    +
    \frac{\partial f}{\partial y}\frac{\partial y}{\partial x}
    = 0,
\end{equation*}
which leads to the desired derivative:
\begin{equation}~\label{eq:implicit-diff}
    \frac{\partial y}{\partial x}
    =
    -\left(\frac{\partial f}{\partial y}\right)^{-1}
    \frac{\partial f}{\partial x}.
\end{equation}
This is a solver-agnostic formula, meaning the implementation of $g$ used to compute $y$ from $x$ (e.g., direct solver, iterative method, non-linear optimization) does not affect the correctness of Equation~\eqref{eq:implicit-diff}---what matters are the partial derivatives of $f$.

Assuming $y$ is fed into a scalar-valued energy to be minimized ${\ell : \R^{n} \to \R}$, we can use the adjoint method as an efficient computational procedure for obtaining the derivatives of $\ell$ with respect to $x$.
Namely, it states that the derivative
\begin{equation}~\label{eq:implicit-full}
    \frac{\partial \ell}{\partial x}
    =
    \underbrace{-\frac{\partial \ell}{\partial y}{\left[\frac{\partial f}{\partial y}\right]}^{-1}}_{\mathlarger{\sfrac{\partial \ell}{\partial f}}}\frac{\partial f}{\partial x},
\end{equation}
can be computed by computing $\sfrac{\partial \ell}{\partial f}$ first, before computing the remaining terms.
This is efficient because $\sfrac{\partial \ell}{\partial y} \in \R^{1 \times n}$ always has fewer or equal elements than $\sfrac{\partial f}{\partial x} \in \R^{n \times m}$, and corresponds to standard backward-mode differentiation.

This approach illustrates a broad pattern where the backward pass through a solver is defined mathematically by the implicit relation, not by the implementation details of the forward algorithm.
This viewpoint enables efficient, sparsity-preserving differentiation for a wide class of geometry processing tasks.
Indeed, the adjoint method has been applied to a variety of inverse tasks all throughout computer graphics and geometry processing~\cite{huang2024diff}, albeit its use usually involves tedious and manual computation of the necessary derivative formulas. Our system automates this computation for geometric applications.

\begin{figure*}[th!]
  \centering
  \includegraphics[width=\linewidth]{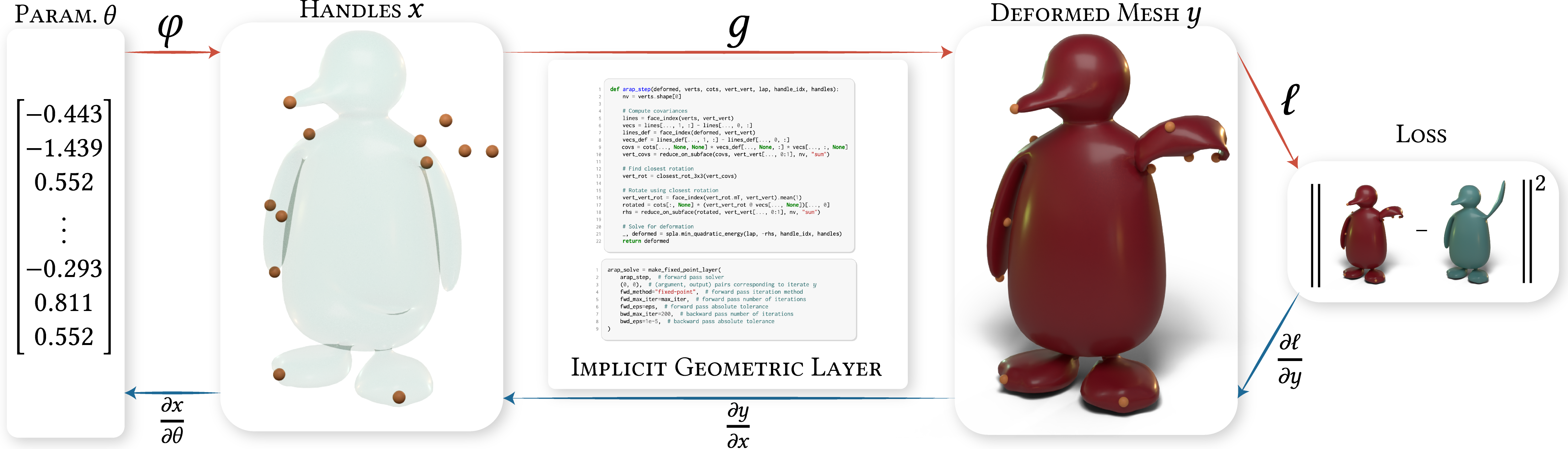}
  \Description{A flowchart describing our method demonstrated on a penguin mesh.}
  \caption{
  A typical pipeline in \texttt{iskra}: the user maps a parameter vector $\theta$ to mesh quantities $x$, in this case, control handles. Next the user defines an \emph{implicit geometric layer}, which solves a forward (inner) geometric problem, such as ARAP deformation; \texttt{iskra} automatically provides gradients in the process.
  Finally, the user uses the solution to the inner problem in an outer problem with loss function $\ell$.
}\label{fig:explainer}
\end{figure*}

\section{Programming Model}\label{sec:prog_model}

In \texttt{iskra}, we consider optimization problems of the form
\begin{equation}
    \min_{\theta \in \mathbb{R}^p} \; \ell\bigl(g(\varphi(\theta))\bigr),
    \label{eq:opt_prob}
\end{equation}
where $\varphi$ maps \emph{outer} optimization variables $\theta \in \R^p$ to quantities defined over a simplicial mesh; $g$ denotes a \emph{geometric solver} that maps these quantities to a solution $y$ implicitly defined by a system of equations on the mesh (e.g., a linear solve, a fixed-point iteration); and $\ell$ is an objective defined on the output of $g$. Both $\varphi$ and $\ell$ are composed from arbitrary differentiable operations, including not only geometry processing operations, but also neural network components.
As a concrete example, inputs to $g$ could include vertex positions, boundary conditions, and finite element operators and the corresponding output could be the geodesic distance function.

From a user's perspective, expressing an inverse geometry problem in \texttt{iskra} follows the straightforward three-step pattern outlined in the next three sections.

\subsection{Specify Parameters and Geometry}\label{ss:specify}

The user first specifies the mesh and the \emph{outer} optimization variables $\theta$.
In code, these are ordinary differentiable PyTorch tensors and can be anything from mesh data to neural network parameters.
The user provides the map $\theta \mapsto \varphi(\theta)$, where $\varphi(\theta)$ is, e.g., a geometric quantity defined on mesh elements or a sparse matrix.
Constructing geometric quantities is straightforward as \texttt{iskra} provides both a low-level scatter-gather interface for working with mesh-structured data (\S\ref{sec:geo_module}), as well as high-level primitives, such as element-wise quantities discrete differential operators (\S\ref{sec:implicit_module}).

In \texttt{iskra}, mesh topology consists of sets of faces each of which is assigned a unique index and a canonical orientation.
We notate the set of $d$-dimensional faces, i.e., those containing $(d+1)$ vertices, as $\mathbb F^d$.
These are $\CS, \TS, \ES, \VS \subset \N$ for tetrahedra, triangles, edges, and vertices.
Their relationships are defined via tensors such as $\EV \in \VS^{|\ES| \times 2}$ or $\VV \in \VS^{2|\ES| \times 2}$; for more details see \S\ref{sec:gather_impl}.
A mesh's geometry is most often defined in terms of vertex positions $\vec v \in \R^{|\VS| \times 3}$.
More generally, each mesh element can be equipped with an attribute, geometry is only a special case.

Continuing with our introductory example of ARAP deformation~\cite{Igarashi:ARAP,Sorkine:ARAP}, we assume a triangular mesh and define a set of control handles $\HS \subset \VS$ with positions $\vec h \in \R^{|\HS| \times 3}$.
As a reminder, our goal is to optimize the handle positions such that the deformed mesh best approximates some target mesh with vertices $\vec t$. 
See Figure~\ref{fig:explainer} for a visual illustration.
 
In \texttt{iskra}, loading meshes and obtaining $\vec t$ and $\vec v$ can be carried out as follows:
\begin{minted}{python}
mesh, _ = Mesh.from_path(mesh_path, dtype=dtype, device=device)
verts = mesh.geom.vertices  # |\ccomment{$\vec v$}|

target_mesh, _ = Mesh.from_path(mesh_path, dtype=dtype, device=device)
target_verts = target_mesh.geom.vertices  # |\ccomment{$\vec t$}|
\end{minted}
We load the handles and initialize their positions to the corresponding vertex positions in the undeformed mesh.
Next we define them as optimization variables $\theta$ following standard PyTorch practice, by marking them as differentiable and passing them to our optimizer of choice, in this case (stochastic) gradient descent (SGD):
\begin{minted}{python}
handle_idx = load_handles(handles_path)  # |\ccomment{$\HS$}|
handles = verts[handle_idx]  # |\ccomment{$\vec h$}|
handles.requires_grad_(True)  # differentiable
optimizer = SGD([handles], lr=lr)  # make |\ccomment{$\vec h$}| into |\ccomment{$\theta$}|
\end{minted}
In this simple example, $\varphi$ is the identity map.%
\subsection{Define Implicit Geometric Layer}\label{ss:defimplicit}

The next step for the user is defining the solver $y \leftarrow g(x)$. To support our goal of versatility, our system is intentionally general and unopinionated at this level. Once such a core is in place, more specialized convenience utilities can be layered on top. At the lowest interface level, users can define a differentiable $g$ by specifying 
\begin{enumerate}
    \item a method for computing $g$ during the forward pass,
    \item an implicit relationship $f$ between the inputs and outputs of $g$, used to automatically create $g$'s backward pass.
\end{enumerate}
More specifically, for a given $g$, the user specifies a corresponding differentiable implicit relation $f(x; y) = 0$ as described in \S\ref{sec:background}, giving $g$ the name \emph{implicit geometric layer}.

This separation of concerns allows $g$ to be an arbitrary function, including calls to external low-level libraries, such as the optimized sparse linear algebra routines like cuDSS~\cite{NVIDIA:2025:cudss} and Cholmod~\cite{Chen:2008:A8C}.
The user implements the implicit relation $f$ using a combination of standard PyTorch routines and \texttt{iskra}'s geometry processing features, making it so the system can use automatic differentiation to assemble an adjoint method backward pass.
At the lowest level, \texttt{iskra} exposes a minimal interface in which users directly specify the forward solver and its corresponding implicit relation. \texttt{iskra} also supports other common patterns for defining $g$, including linear solves, spectral problems, and fixed-point iterations that arise in geometric processing. 

For ARAP specifically, in addition to our optimization variables, we need a set of auxiliary variables, namely the cotan weights $\vec c \in \R^{|\ES|}$ which we can map onto each oriented vertex pair, thus $\hat{\vec c} \in \R^{2|\ES|}$, and assemble into the Laplacian matrix $L \in \R^{|\VS| \times |\VS|}$:
\begin{minted}{python}
faces = mesh.topo.faces  # |\ccomment{$\FS$}|
vert_vert = vertex_adjacency(faces)  # |\ccomment{$\VV$}|

weights = cotan_weights(verts, faces, clamp_min=1e-5)  # |\ccomment{$\vec c$}|
vert_vert_weights = edge_to_vertex_adjacency(weights)  # |\ccomment{$\hat{\vec c}$}|
lap = laplacian_from_weights(weights, faces)  # |\ccomment{$L$}|
\end{minted}
In ARAP, it is common to pre-factor the Laplacian for efficiency.
All of our ARAP results were generated with this optimization, but we choose to omit it here for didactic purposes.

Next, we define $f$. In \texttt{iskra}, we can use ARAP's iterative optimization loop to define the implicit relationship:
\begin{minted}{python}
def arap_step(deformed, verts, cots, vert_vert, lap, handle_idx, handles):
    nv = verts.shape[0]

    # Compute covariances
    lines = face_index(verts, vert_vert)
    vecs = lines[..., 1, :] - lines[..., 0, :]
    lines_def = face_index(deformed, vert_vert)
    vecs_def = lines_def[..., 1, :] - lines_def[..., 0, :]
    covs = cots[..., None, None] * vecs_def[..., None, :] * vecs[..., :, None]
    vert_covs = reduce_on_subface(covs, vert_vert[..., 0:1], nv, "sum")

    # Find closest rotation
    vert_rot = closest_rot_3x3(vert_covs)

    # Rotate using closest rotation
    vert_vert_rot = face_index(vert_rot.mT, vert_vert).mean(1)
    rotated = cots[:, None] * (vert_vert_rot @ vecs[..., None])[..., 0]
    rhs = reduce_on_subface(rotated, vert_vert[..., 0:1], nv, "sum")

    # Solve for deformation
    _, deformed = spla.min_quadratic_energy(lap, -rhs, handle_idx, handles)
    return deformed
\end{minted}
We can convert this into an implicit relationship between the relevant optimization variables with a simple wrapper:%
\begin{minted}{python}
def f(handles, deformed):
    return deformed - arap_step(deformed, |\color{gray!90}{\ldots}|, handles)
\end{minted}

Finally, we are ready to assemble $g$.
For now, we assume access to an ARAP solver \mintinline{python}{arap_solver(handles)}, be it one we created using \mintinline{python}{arap_step}, or one of the many accelerated methods that optimize objectives like ARAP~\cite{bourquat2022hierarchical,manson2011hierarchical,peng2018anderson}.
In forming $g$, we must specify which arguments of $f$ serve as $x$ and which as $y$, i.e., which arguments should be treated as functions of which other arguments.
In this case, we wish to find the derivative of \mintinline{python}{deformed}, in position $1$, w.r.t. \mintinline{python}{handles}, in position $0$.
While our system allows the user to specify which iterative solver to use for the linear system in the adjoint formula, we rely here on the default GMRES~\cite{Saad:1986:GAG} solver:
\begin{minted}{python}
arap_solve = make_adjoint_layer(
    arap_solver,  # forward solver
    f,  # implicit relationship
    in_args=0,  # which arguments correspond to |$x$|
    out_args=1,   # which arguments correspond to |$y$|
    bwd_max_iter=200,  # number of adjoint solver steps
    bwd_rel_tol=1e-5,  # adjoint solver relative tolerance
)
\end{minted}
This is intentionally a very general formulation meant to cover the broadest possible range of inverse problems one might wish to solve---as long as there is a known implicit relationship between the inputs and outputs of the solver and $\frac{\partial f}{\partial y}$ is invertible at a given $x$, our system can assemble a backwards pass.

However, we can offer greater usability in special cases by leveraging the fact that $g$ and $f$ are oftentimes more closely related.
For example, $g$ is often best implemented as a fixed-point iteration $y \leftarrow \widetilde f(x; y)$, e.g., a contraction map converging to $y$ (see \S\ref{ss:arap}, \S\ref{ss:geodesics}).
Our system also offers a higher-level interface for this specific case, automatically generating the forward pass from $\widetilde f$, and generating the backward pass through the fact that this fixed-point problem is isomorphic to $f(x; y) = \widetilde f(x; y) - y = 0$.
For our running example, \texttt{iskra} allows us to create a fixed-point solver from \mintinline{python}{arap_step} by specifying which arguments and outputs correspond to $y$:
\begin{minted}{python}
arap_solve = make_fixed_point_layer(
    arap_step,  # forward pass solver
    (0, 0),  # (argument, output) pairs corresponding to iterate |$y$|
    fwd_method="fixed-point",  # forward pass iteration method
    fwd_max_iter=max_iter,  # number of fixed-point iterations
    fwd_abs_tol=abs_tol,  # forward pass absolute tolerance
    bwd_max_iter=200,  # number of adjoint solver steps
    bwd_rel_tol=1e-5,  # adjoint solver relative tolerance
)
\end{minted}
Regardless of exact interface, the user writes their code by composing \texttt{iskra}'s sparse mesh-specific operators and \texttt{iskra} attaches to each forward pass solver a corresponding implicit backward rule, thus avoiding naively differentiating through internal iterations.

\subsection{Compose a Geometry Objective}

Having defined $\theta$, $\varphi$, and our implicit geometric layer $g$, the user specifies a scalar objective $\ell$ %
as a composition of differentiable operations.
Typical choices can include standard geometry energies, e.g., symmetric Dirichlet, various smoothness energies, geodesic distance penalties.
Our system is flexible enough to allow the user to incorporate more complex machine learning components, such as neural network-based energies.

In our specific scenario, we wish to match the deformed vertices produced by $g$ to the target vertices $\vec t$,
\begin{equation}\label{eq:invarapprob}
    \ell(\vec t, \theta) = \frac{1}{|\VS|} \lVert \vec t - g(\varphi(\theta)) \rVert^2_2,
\end{equation}
where we take the $L_2$ norm instead of the mass-matrix norm to simplify exposition.
From the user's point of view, this example objective consists only of ordinary code written in PyTorch where the only difference is that some tensors result from implicit geometric layers:
\begin{minted}{python}
optimizer.zero_grad()
deformed = arap_solve(
    verts, vert_vert_weights, vert_vert, lap, handle_idx, handles
)
loss = ((deformed - target_verts) ** 2).sum(-1).mean()
loss.backward()
\end{minted}
For a more complex version of the problem involving real-world motion capture data~\cite{AMASS}, refer to \S\ref{ss:arap} and Fig.~\ref{fig:amass}.

Having already chosen the outer optimizer (e.g., SGD, Adam), the user can perform gradient-based optimization over $\theta$ by iteratively computing the loss and stepping with the optimizer.
The outer optimization loop is therefore written entirely in standard automatic differentiation style, while the complexity of differentiating through solvers is handled internally by \texttt{iskra}.

In our example, we note that the \mintinline{python}{arap_step} function hides within it two nested adjoint problems, namely the sparse linear solve within \mintinline{python}{spla.min_quadratic_energy} and the \mintinline{python}{polar3x3} matrix decomposition. This is permissible in \texttt{iskra}, as many geometry pipelines solve smaller optimizations as part of a larger problem.

\subsection{Capabilities}\label{sec:capabilities}

As we have seen, \texttt{iskra} uses implicit differentiation of $f(x; y) = 0$ to obtain gradients $\partial \ell / \partial \theta$ and integrates them to the outer optimizer, regardless of the optimization type.
We emphasize the breadth of the set of problems covered by our formulation.
Specifically, some of the problems that fall under this umbrella are linear systems (\S\ref{ss:mcf}), eigenproblems (\S\ref{ss:scp}), and problems solvable via local-global solvers (\S\ref{ss:arap}) or ADMM iterations (\S\ref{ss:geodesics}).
Moreover, our system also allows for flexibility in how we solve $g$'s backward pass.
This has significant impact on efficiency compared to other approaches, both in terms of memory requirements and speed (see \S\ref{sec:apps}).

To amplify usability and efficiency of our system, %
we include special functionality for two pervasive operations that have structure beyond the generic case.  In particular, we provide bespoke layers that implement two strongly-structured forms of $f$ that appear in nearly all geometry processing algorithms:
\begin{itemize}
    \item \textbf{Linear implicit layers}, where the forward pass solves 
    \begin{equation}    
        A\,\vec y = \vec b,
    \end{equation}
    with the accompanying implicit relationship:
    \begin{equation}
        \vec f(A, \vec b; \vec y) = A\,\vec y - \vec b = 0.
    \end{equation}
    See \S\ref{sec:implicit_module} for details.
    \item \textbf{Spectral implicit layers}, where the forward pass solves 
    \begin{equation}
    \begin{aligned}
        & AU = UB\diag(\vec \lambda), \\
        & U^\tpose B  U = I,
    \end{aligned}
    \end{equation}
    where the columns of $U \in \R^{n \times k}$ and entries in $\vec \lambda \in \R^k$ represent eigenvectors and eigenvalues of the generalized eigenvalue problem defined by $A \in \R^{n \times n}$, and $B \in \R^{n \times n}$.
    The eigenvector problem is special not only because of its ubiquity in geometry processing, but also because it admits many ways of differentiating through its solutions. %
    Therefore, our system allows the user to switch between these options. We include a thorough study of the different trade-offs (\S\ref{ss:scp}).
    See \S\ref{sec:implicit_module} for examples. 
\end{itemize}

\section{\texttt{iskra} System Design}\label{sec:iskra}

\texttt{iskra} is organized around two tightly coupled modules: a \emph{differentiable geometry module} and an \emph{implicit-function module}.
These intertwined modules work together to enable inverting geometry processing problems.
The \emph{differentiable geometry module} (\S\ref{sec:geo_module}) is responsible for representing meshes, mesh relations, local geometric operators, and offers additional support for irregular data and geometry-specific quantities like (dual) quaternions.
The \emph{implicit-function module} (\S\ref{sec:implicit_module}) is responsible for differentiating through solvers using the adjoint method, as discussed previously (\S\ref{sec:background}, \S\ref{sec:prog_model}).

The design of the geometry module serves as a foundation for the implicit-function module.
Supporting a broad range of existing geometry algorithms requires us to give full control to the user; typically, such low-level control would imply a higher implementation burden, standing in an apparent contradiction with our other stated design goal of usability.
This tension informs the design of our geometry module.
It aims to be as compact and efficient as possible as a way of alleviating the user effort for expressing existing algorithms within the implicit-function module.
Thus, the representations of mesh relations, operators, and data movement are chosen to preserve computational sparsity central to geometric workloads during implicit differentiation, all while integrating naturally with the broader tensor-based ecosystem.
Crucially, neither is sufficient by itself: their interaction enables \texttt{iskra} to achieve the stated design goals of versatility, usability, interoperability, and efficiency.

\subsection{Geometry Module}\label{sec:geo_module}

The geometry module of \texttt{iskra} is responsible for representing meshes, mesh-based data, and constructing local geometric operators.
Rather than introducing a specialized domain-specific language or execution model~\cite{mahmoud2021rxmesh, yu2023taichi}, which sacrifice flexibility and interoperability for raw speed, \texttt{iskra}'s tensor-centric design necessitates a different approach.
While sharing the tensor abstraction with other systems~\cite{libigl,gptoolbox,gpytoolbox}, \texttt{iskra} builds upon them by systematizing operations on mesh data represented this way.
Specifically, inspired by~\citet{yu2023taichi}, \texttt{iskra} expresses geometry processing as a combination of \emph{gather} and \emph{scatter-reduce} operations defined on mesh topology.
A key realization in our design is that the scatter-gather paradigm maps quite naturally to the tensor abstraction, even when it comes to mesh attribute data.

\texttt{iskra}'s design thinks of mesh faces in a hierarchy. For example, in a one-dimensional polyline mesh, line-segments (1D faces) are at the highest-level and contain within them vertices (0D faces).
Therefore, we say that a vertex is a \emph{subface} of the line-segment. We use this idea to describe the relation between edges (1D faces) and triangles (2D faces).
Our system's topology interface allows us to extract all topology relationships commonly found in geometry processing, including extracting lower levels of the hierarchy from the higher ones, taking into account orientations when needed.

Mesh-based operations are expressed using explicit \emph{gather} and \emph{scatter-reduce} transformations to move across this hierarchy.
Conceptually, these operations follow two complementary patterns:
\begin{itemize}
    \item \emph{Gather} operations read attributes defined on a subface of a mesh element and collect them all on the face.
    \item \emph{Scatter-reduce} operations distribute and accumulate attributes defined on a face to its subfaces.
\end{itemize}
Thus, gather operations move data up the hierarchy, whereas scatter-reduce operations move data down the hierarchy.
In combination, these patterns can express common local geometry processing tasks, such as assembling discrete differential operators, computing per-element energies, or accumulating forces. 
These operations are element-agnostic, i.e., the same abstraction applies uniformly to vertices, edges, triangles, or tetrahedra. 
In \texttt{iskra}, they are implemented as tensor operations, allowing the same code to run on CPU or GPU and to participate in automatic differentiation.

A simple example demonstrating this approach is computing normals on a 1D mesh embedded in 2D.
First, using the edge-to-vertex relationship $\EV$, we gather all of the 2D vertex positions into a $2 \times 2$ tensor on each edge.
Next, we compute the per-edge normal vectors.
Lastly, we scatter an edge's normal vector onto the vertices belonging to that edge before finally normalizing. We can express this operation in \texttt{iskra} as:
\begin{minted}{python}
# Gather vertex positions onto edges using |$\EV$|:
lines = face_index(verts, edge_vert)

# Compute per-edge normal:
vecs = lines[..., 1, :] - lines[..., 0, :]
orth = torch.broadcast_to(torch.tensor([1.0, -1.0]), vecs.shape)
face_normals = vecs[..., (1, 0)] * orth_vector
 
# Scatter-reduce from edges to vertices using |$\EV$|:
normals = reduce_on_subface(face_normals, edge_vert, verts.shape[0], "sum")
normals = normalize(normals, dim=-1)
\end{minted}

This simple example demonstrates the basic programming model of our system, albeit more complicated operations are possible.
For example, by default, the system assumes all attributes are vector-valued and decides on the scattering strategy based on that.
In some situations, ambiguities arise, e.g., a $|\TS| \times 3 \times 3$ tensor could either represent a $3\times3$ matrix per face, or a $3$-vector per corner of a face.
Our system allows the user to specify the dimensionality of the element attribute and decides the scattering strategy based on this.

In addition, due a lack of fully-fledged support for sparsity in PyTorch, we provide the user with a custom \mintinline{python}|SparseTensor|, implemented using the \emph{coordinate} sparse format, offering standard utilities common to tensor based frameworks such as indexing, slicing, reshaping, concatenation, as well as elementwise and binary operations.

\paragraph{Discussion.}
By representing mesh relations through explicit sparse gather and scatter operations, \texttt{iskra} simultaneously achieves three design goals.
First, it preserves the computational sparsity inherent in discrete geometry operations.
Second, it aligns mesh-based computation with the dataflow model expected by modern automatic differentiation systems.
Third, it provides a uniform interface that supports a wide range of geometric operators without introducing a separate execution or programming model.
Our design inherits trade-offs common to tensor-based frameworks, namely it favors simplicity, composability, and interoperability in lieu of  peak mesh traversal performance or compressed memory storage.
In practice, we expect our system to integrate with other resource heavy workloads, including, e.g., machine learning and inverse rendering.
Due to this, we do not expect our system to be the bottleneck and leave for future work to explore low-level acceleration strategies.

\subsection{Implicit-Function Module}\label{sec:implicit_module}

Having described the foundation that allows our system to remain efficient even when faced with sparse data flows of geometry processing, we now turn to the implicit function module. 
The difficulty lies in differentiating through the solver $y \leftarrow g(x)$ from Equation~\eqref{eq:opt_prob}.
Towards this goal, \texttt{iskra}'s \emph{implicit-function module} provides a uniform mechanism for evaluating and differentiating through a broad set of solvers $g$.
This module is responsible for both the forward evaluation of $g$ and the backward propagation of its gradients.

Geometry processing problems give rise to a wide variety of problems and solvers, e.g., linear systems, generalized eigenproblems, alternating local-global schemes, or problem-specific ADMM instances.
Rather than designing the system around a single class of optimization problems, the implicit-function module is formulated as generally as possible to support this diversity.
At the lowest level, if $g$ solves a known optimization objective the supplied implicit relationship $f$ is used to construct the gradient of the solution with respect to $g$'s inputs.
As we will see, in many cases a higher-level interface is possible, e.g., it can be sufficient to specify one step of a fixed-point iteration scheme.
Succinctly, our design objective for this module is to accommodate the need for wide diversity of problem-specific solvers, while offering a sufficiently pliable, yet ergonomic experience to geometry processing experts.

To achieve this level of generality, \texttt{iskra} adopts an \emph{imperative} formulation for its implicit layers.
As covered previously, to generate the backward pass, the user supplies an $f$ that they themselves wrote using functionality from \texttt{iskra}'s geometry module.
The system does not require the user to specify the objective of $g$ in a declarative form, nor to restrict problems to a specific optimization template.
This choice contrasts with declarative approaches, such as \textsc{CVXPYLayers}~\cite{agrawal2019cvxpylayers}, $\nabla$-Prox~\cite{lai2023deltaprox}, or Theseus~\cite{pineda2022theseus}, which demand problems be within the narrow class of either convex, proximal, or non-linear least squares to derive gradients automatically.
Using them involves reformulating optimization problems into the specific declerative form they dictate, which can come with its own challenges (\S\ref{sec:apps}).

Such approaches are not well suited to geometry processing workloads which routinely involve highly nonlinear or non-convex energies (\S\ref{ss:arap}, \S\ref{ss:scp}), or involve problem with known exploitable structures (\S\ref{ss:geodesics}).
Taken together, the scope of problems in geometry processing does not admit a single canonical declarative representation nor does it allow for a one-size-fits-all set of solvers.
By adopting an imperative formulation, \texttt{iskra} allows each solver to be implemented using the most appropriate numerical method while enabling implicit differentiation via automatically derived adjoints.

For each implicit-function layer, the forward pass computes $y$ by solving the implicit equation using an arbitrary, possibly black-box solver $g$ while the backward pass differentiates through its solution using implicit differentiation.
This separation allows users to rely on fast specialized solvers even if their implementations are not differentiable.
At a high level, the this module consists of:
\begin{itemize}
    \item an interface for specifying and invoking geometry-aware solvers to compute $y$ from $x$;
    \item an interface that allows the user to imperatively specify the implicit relationship $f(x; y) = 0$,
    \item a set of problem class-specific routines that exploit the sparsity and structures common to that class.
\end{itemize}
While our applications in \S\ref{sec:apps} follow the structure in Equation~\eqref{eq:opt_prob}, in practice our system allows users to stack and nest implicit geometric layers and interleave them with other differentiable workloads.

The core insight enabling our design is that, unlike the forward pass, which can be solved in arbitrary ways depending on the problem, the backward pass is guaranteed to be a sparse linear system.
After all, the Jacobian $\sfrac{\partial f}{\partial y}$ that we must invert is a linear approximation of a function at a given point.
This means that, regardless of the complexity of $g$ or $f$, the backward pass is effectively a (sparse) linear solver (see \S\ref{sec:implicit_diff}).
In all but the simplest cases, the Jacobian at any given $x$ is likely an enormous matrix; even with sparsity accounted, using a direct solver would likely be prohibitive for large problems.
Instead, we rely on an iterative method, allowing the user to use a customized solver if needed; for more information, see \S\ref{sec:adjoint_impl}.

Since our loss $\ell$ is a scalar-valued function, we can leverage the adjoint method.
Practically, this means that given a function $f(x; y)$, our system first constructs two functions that evaluate \emph{vector-Jacobian products}; namely, one that right-multiplies its inputs with $\sfrac{\partial f}{\partial y}$, and another that right-multiplies with $\sfrac{\partial f}{\partial x}$.
Then we apply the adjoint method: first we compute $\sfrac{\partial \ell}{\partial f}$ by using an iterative solver that repeatedly applies $\sfrac{\partial f}{\partial y}^\tpose$ to $\sfrac{\partial \ell}{\partial y}^\tpose$, before incorporating $\sfrac{\partial f}{\partial x}$ to compute the full expression.
This approach to differentiation is used regardless of whether the user relies on the fixed-point interface or the more general low-level one.

The fixed-point interface contributes to the usability of our system, allowing users to solve very broad classes of problems.
For example, in addition to the local-global solver in our running ARAP example (\S\ref{sec:prog_model}), our system is demonstrably able to differentiate through problems solvable using ADMM iterations (\S\ref{ss:geodesics}). As long as a fixed-point iteration exists, our system can differentiate through it.
Specific to this interface is that we allow the user to incorporate additional information into the generation of the forward pass, such as selecting a function's output to serve as the energy objective or specifying tolerances (see \S\ref{sec:adjoint_impl}).

Two problem classes stand out in how common they are in geometry processing and in our ability to provide specialized efficient solutions for computing their derivatives, namely sparse linear systems and sparse eigensolvers.
Besides being ubiquitous in geometry, linear systems are also mathematically special.
They are the only problem class where the Jacobian $\sfrac{\partial f}{\partial y}$  of the backward pass happens to be the same as the forward problem, opening doors to further optimizations.
Eigensystems are special in that they come with a variety of different derivative formulae, each with a different trade-off.
Below we describe iskra's special functionality for these two ubiquitous operations.

\paragraph*{Differentiating linear solvers}
Implementing the adjoint method for a linear system amounts to solving a system with the same matrix as in the forward problem.
Differentiating the implicit relationship ${\vec f(A, \vec b; \vec y) = A\vec y - \vec b = \vec 0}$ with respect to its inputs, we obtain:
\begin{equation*}
    \frac{\partial \vec f}{\partial \vec y} = A,
    \qquad
    \frac{\partial \vec f}{\partial \vec b} = -I,
    \qquad
    \frac{\partial f_i}{\partial A_{jk}} = \delta_{ij} y_k,
\end{equation*}
where $\delta$ is the Kronecker delta.
Incorporating this into the adjoint method from Equation~\eqref{eq:implicit-full}, we see that solving for $\sfrac{\partial \ell}{\partial y}$ involves inverting $A$.
Unlike the general adjoint problem from prior sections, users could exploit knowledge about their specific linear system to accelerate the backward pass---provided the right system affordances.
For example, it should be possible to exploit known matrix structures such as positive semidefiniteness, to reuse precomutation or to rely on a solver that exploits hardware-specific characteristics.
Given how common sparse linear systems are in geometry, staying faithful to our design goals of versatility and efficiency means offering a bespoke solution for this particular case.
In doing this, it is imperative that the system offer good defaults, but also allow the user provide their own sparse solver.
Therefore, we differentiate through the solutions of linear systems by manually implementing the adjoint method, relying on a user-supplied non-differentiable solver when specified. For example:
\begin{minted}{python}
factors = factor_matrix(mat)
def custom_solver(b):
    return solve_with_factors(factors, b)  # not differentiable

x = linear_solve(mat, b, solver_fn=custom_solver)  # differentiable
\end{minted}
We demonstrate how this additional flexibility leads to efficiency gains during both the forward and backward passes (\S\ref{ss:mcf}).

\paragraph*{Differentiating eigenvectors and eigenvalues}
A similarly foundational module of geometry processing is the ability to solve sparse (generalized) eigenvalue problems.
With the same rationale of expanding our system's versatility and efficiency, our system provides additional affordances for eigenproblems.
However, in contrast to linear systems, differentiating through solutions to eigenproblems---specifically through eigenvectors---is less straightforward.
Deriving a closed-form formula for differentiating through a single eigenvector can be done by writing down the equation:
\begin{equation}
\begin{aligned}
    & A\vec u_i = \lambda_i B \vec u_i, \\
    \suchThat\; & \vec u_i^\tpose B \vec u_i = 1,
\end{aligned}
\end{equation}
and assembling the corresponding implicit relationship,
\begin{equation}
\vec f(A, B; \vec{u}_i, \lambda_i) = 
\begin{bmatrix} 
    A \vec u_i - \lambda_i B \vec u_i \\
    \frac{1}{2}(1 - \vec u_i^\tpose B \vec{u}_i)
\end{bmatrix} = 
\begin{bmatrix} 
    \vec 0 \\
    0 
\end{bmatrix},
\end{equation}
where we treat $\vec u_i$ and $\lambda_i$ as dependent on $A \in \R^{n \times n}$ and $B \in \R^{n \times n}$.
Manually computing the required partial derivatives reveals the following expressions:
\begin{equation}\label{eq:eigindividual}
\begin{aligned}
    \frac{\partial \vec f}{\partial (\vec u_i; \lambda_i)} &= \begin{bmatrix}
        A - \lambda_i B & -B \vec{u}_i \\ 
        -\vec u_i^\tpose B & 0 
    \end{bmatrix}\\
    \frac{\partial f_j}{\partial A_{kl}} &= \delta_{jk} (u_i)_l \\
    \frac{\partial f_j}{\partial B_{kl}} &= \begin{cases} 
    -\lambda_i \delta_{jk} (u_i)_l & j \leq n \\
    -\frac{1}{2}(u_i)_k (u_i)_l & j = n + 1
    \end{cases}
\end{aligned}
\end{equation}
On one hand, the system corresponding to $\sfrac{\partial f}{\partial y}$ is sparse, symmetric, and, assuming no repeated eigenvalues, invertible.
On the other, the matrix itself contains both the eigenvalue and eigenvector, meaning that we cannot reuse the same factorization in cases where we need more than one eigenvector.
Thus, if more than one eigenvalue-eigenvector pair is required, our system must loop over each pair and factor and solve a different linear system.

The second approach offered by our system is that of \citet{smirnov2021hodgenet}.
Focusing on derivatives of an eigenvector w.r.t. $A$, further manipulation tells us that they can be computed as follows:
\begin{equation}\label{eq:eigensmirnov}
\begin{aligned}
\frac{\partial \vec u_i}{\partial A_{kl}} = \sum_{m \neq i} \frac{(u_m)_k (u_i)_l}{\lambda_i - \lambda_m} \vec u_m\end{aligned}
\end{equation}
While this compact formula can be implemented using only dense matrix multiplies, it comes at the cost of needing access to all $n$ eigenvalues, something that is most often prohibitively expensive.
To this extent  \citet{smirnov2021hodgenet} introduce an approximation by truncating small eigevalues: if the user requested only the top-$k$ eigenvalues, the method treats the remaining ones as zero.

The remaining approach relies on our system's fixed-point capabilities instead of on hand-derived formulae through a
fixed-point equation based on QR-iterations:
\begin{equation}\label{eq:eigiter}
    f(A, B, U) = Q({(A - \sigma I)}^{-1} U) - U = 0,
\end{equation}
where $Q$ orthonormalizes its inputs w.r.t.\ $B$ and chooses a canonical sign for each column, and $\sigma$ is the \emph{shift-invert} parameter.

Ultimately, the best approach depends on the use-case.
In addition to offering such a broad spectrum of approaches, we include a thorough investigation into the performance characteristics of each, including speed, memory consumption, and relationship to the size of $A$ and the number of requested eigenvectors $k$ in \S\ref{ss:scp}.
\section{Implementation Details}\label{sec:details}

Our framework offers numerous utilities related to mesh processing, including various topological operations, point-to-simplex projections and distances, barycentric coordinate computation, normals and (hyper-)volume computation, quaternion and dual quaternion support, and simplex quality metrics.
We also offer fully fledged discrete exterior calculus \cite{Hirani2003, desbrun2006dec, crane2013deccourse} and finite element toolboxes, representing the discrete operators as sparse matrices.
The entire system is differentiable as it is implemented end-to-end in PyTorch.

In this section we focus on the low-level implementation details, namely those of the scatter-gather framework, our implementation of the adjoint method, as well as specialized optimizers required to make differentiable geometry processing integrate well with other optimization pipelines.

\subsection{Gather/Scatter}\label{sec:gather_impl}

Our scatter-gather framework is suited specifically to meshes, requiring a purpose-built topology toolbox.
Given a simplicial mesh, our system can construct the entire face hierarchy upon user request.
Each level of the hierarchy consists of three dense tensors connecting it to the level below it.
As explained in \S\ref{sec:prog_model}~and~\ref{sec:geo_module}, we assign each $d$-dimensional face a unique index in $\mathbb F^d$ and connect it to $(d-1)$-dimensional faces.
Bookkeeping for this is via a set of $3$ dense tensors per level of the hierarchy: \emph{face-vertex}, \emph{face-subface}, and \emph{subface-orientation} tensors.
For example, for a set of tetrahedra $\CS$, we compute the \emph{face-vertex} tensor 
$\Cv \in \VS^{|\CS| \times 4}$, the \emph{face-subface} tensor as $\CT \in \TS^{|\CS| \times 4}$ where $\CT_{ij} \in \TS$ is the $j$\textsuperscript{th} triangle in the $i$\textsuperscript{th} tetrahedron, and the \emph{subface-orientations} as ${\{-1, 1\}}^{|\CT| \times 4}$, defined relative to the canonical orientations of triangles in $\Tv$.
Moreoever, the triangle $\CT_{ij}$ is guaranteed to be opposite vertex $\Cv_{ij}$: note that this means that, e.g., $\EV$ and $\Ev$ are not the same tensors.
In addition, we provide the connections $\textsc{C}_E$ and vertex adjacencies $\VV \in \VS^{2|\ES| \times 2}$ for convenience.
In addition to supporting open and non-manifold meshes, this representation could also be used to represent meshes consisting of mixed elements.

Gathering data from subfaces onto faces is done via the \mintinline{python}{face_index} function which accepts a tensor of $k$-dimensional values, a \emph{face-subface} connection and returns a tensor of the same shape as the face-subface tensor with additional dimensions for the gathered data.
Scattering from faces to subfaces is done via the \mintinline{python}{reduce_on_subface} function, accepting a set of values defined on the faces and a face-subface connection.
We implement both using PyTorch's built-in scatter-gather functions.

\subsection{Adjoint Method Implementation}\label{sec:adjoint_impl}

As discussed in Section~\ref{sec:implicit_diff}, our system constructs vector-Jacobian products from $f$ using PyTorch's standard automatic differentiation capabilities(i.e., functionality from \mintinline{python}{torch.autograd}).
Given a function $f$ and the \mintinline{python}{out_args} parameter, we construct a function that computes vector-Jacobian products with $\sfrac{\partial f}{\partial y}$.
We pass this function to GMRES, which repeatedly applies it to solve for $\sfrac{\partial \ell}{\partial y}$.
Then we use the \mintinline{python}{in_args} parameter to construct the second vector-Jacobian product.
It is up to the user to ensure that $f$ uniquely determines $y$ in terms of $x$ and that $\sfrac{\partial f}{\partial y}$ is invertible at $x$.
Since we are dealing with an intentionally broad formulation, we assume no structure on the Jacobian $\sfrac{\partial f}{\partial y}$, leaving us to rely on the very general GMRES~\cite{Saad:1986:GAG} linear solver for the adjoint equations.
While we have not used these features for our experiments, we allow the user to pass in a custom function instead of our GMRES implementation, technically allowing for preconditioning or carefully chosen initializers.

We provide the ability to specify absolute and relative tolerances to GMRES, as well as the maximum number of iterations.
If \mintinline{python}{max_iter} is reached and the tolerances are unmet, the system prints a warning.
Setting \mintinline{python}{verbose=True} exposes debugging information, including absolute and relative solver errors.
Alternatively, the user can pass \mintinline{python}{strict=True} as an argument to GMRES to raise an error in this scenario.
See \S\ref{s:discussion} for discussion of GMRES stability.
We did not aim to optimize our implementation of GMRES.
In particular, the memory required by GMRES scales linearly with the number of iterations performed.
We leave optimizing memory usage to future work.

In our fixed-point implementation, we provide additional utilities.
The user can specify absolute and relative tolerances for the fixed-point solver.
By default, these are compared to the difference between consecutive iterates; alternatively, the user can designate one of the iteration function's outputs as the optimization objective.
More information can be exposed via \mintinline{python}{verbose=True}, and if the user prefers certainty, they can set \mintinline{python}{max_iter=None}, which enforces the tolerances be fully satisfied.
We use fixed-point tolerances in \S\ref{ss:geodesics}.

We offer two sparse linear solvers to the user by default: CUDSS \cite{NVIDIA:2025:cudss} and CHOLMOD~\cite{Chen:2008:A8C}.
Our default choice is to rely on CHOLMOD on the CPU and CUDSS on the GPU.
Both solvers allow reusing parts of the factorization process, a common pattern in geometry research.
For the eigensolver, we rely on ARPACK~\cite{Lehoucq:1998:AUG} via SciPy~\cite{2020SciPy-NMeth} on the CPU and custom power-iteration code on the GPU.

Lastly, we include two specialized optimizers to the user that are tailored to geometry applications, a simple line-search procedure, and a so-called Sobolev $H^1$ preconditioner that allows us to use gradient-based optimization on vertex data~\cite{neuberger1985steepest,karatson2005sobolev,osher2022laplacian}.
We use these optimizers in Sections~\ref{ss:mcf}~and~\ref{ss:geodesics}.
\begin{figure*}[!ht]
  \centering
  \includegraphics[width=\linewidth]{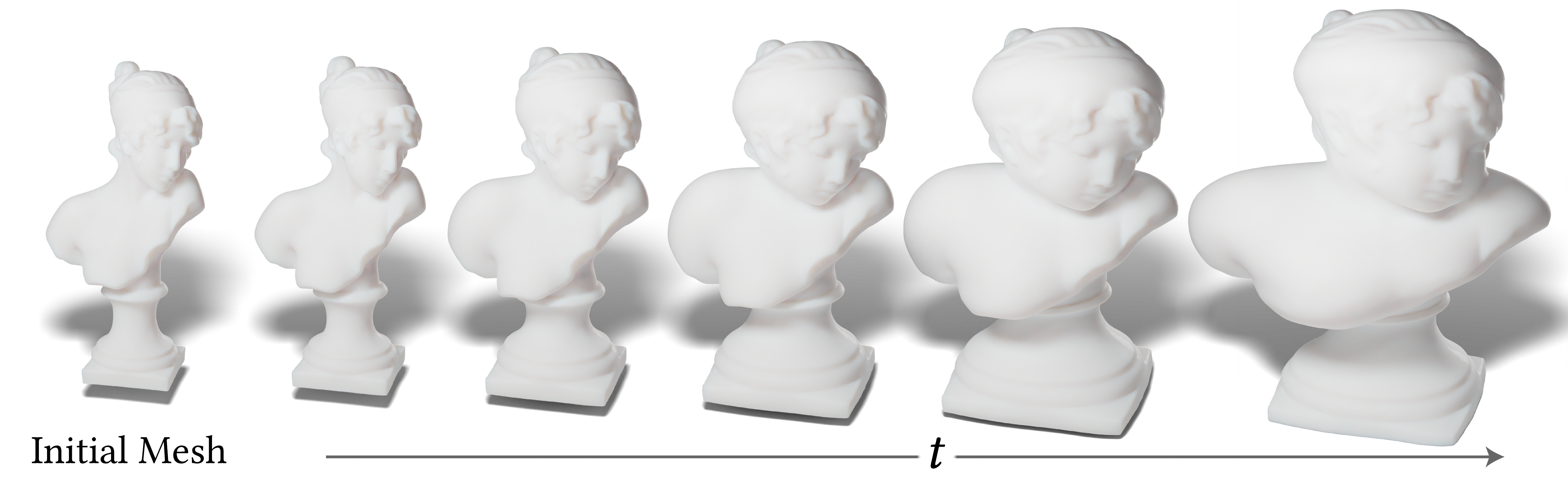}
  \Description{A statue of Sappho. Left initial undeformed mesh. Right, progressively more inflated statues of the same mesh.}
  \caption{Mesh inflation via inverse mean curvature flow. The \textsc{Sappho} mesh is inflated by treating it as the solution to the mean curvature flow at some time $t$ and optimizing for the initial condition that led to that solution. Larger $t$ results in more inflation.}\label{fig:invmcf}
\end{figure*}

\section{Applications}\label{sec:apps}

In this section, we demonstrate how our system achieves its stated goals by implementing a number of inverse geometry processing problems and comparing them with alternative approaches.
We demonstrate versatility by showing its application to four different geometry processing problems, each with a different set of requirements.
We illustrate usability and interoperability by comparing our code to the code needed to solve these problems in declarative systems.
Moreover, our system is interoperable with machine learning pipelines, as evidenced through its integration with a neural field~\cite{xie2022neural} in \S\ref{ss:scp}.
Lastly, for each problem, we include a thorough timing comparison with an alternative system compatible with that problem class.

Our experiments aim to demonstrate creative applications enabled by our framework, in addition to validation and benchmarking.
They show how our system enables efficient implementation of unexplored classes of inverse geometric algorithms.
For existing practical applications of inverse geometry processing, see Section~\ref{ss:diffgp}.

First, we demonstrate how to invert a simple fairing problem~\cite{desbrun2023fairing} (\S\ref{ss:mcf}).
Doing so repeatedly within an optimization loop necessitates fast  solving of sparse linear systems; as we will see, the additional flexibility of our linear system interface enables the user to exploiting problem-specific structures to extract greater performance (Figure~\ref{fig:invmcf_time}).
Next, we show how our system enables differentiating through the solution of a parameterization algorithm formulated as an eigenproblem~\cite{Mullen:2008:SCP}, offering a thorough comparison of various formulations.
We then demonstrate our system's performance on the nonlinear, non-convex ARAP problem~\cite{Sorkine:ARAP} from \S\ref{sec:prog_model}, including an example of how \texttt{iskra} could be used to fit a deformation of a registered mesh to real motion capture data (Fig.~\ref{fig:amass}).
Finally, we compare our system's performance on the quadratic programming \emph{regularized geodesic distance problem} (RGD)~\cite{edelstein2023rgd}.
ARAP and RGD both come with bespoke solvers proposed by the geometry processing community, namely ARAP's local-global solver and RGD's custom ADMM scheme, each of which outperforms general-purpose optimization systems for its problem class.
As we will see, using these schemes via our fixed-point interface serves the dual purpose of performance benefits and allowing practitioners to differentiate through these algorithms without needing to reformulate them.

\subsection{Inverse Mean Curvature Flow}\label{ss:mcf}

Our system allows us to differentiate through the fairing problem proposed by~\citet{desbrun2023fairing}.
If applying mean curvature flow to a mesh thins it out, then the inverse problem would recover an inflated version of the mesh.
Differentiating through mean curvature flow is a highly ill-posed problem and using it to inflate a mesh requires additional regularization, in our case, in the form of Sobolev $H^1$ preconditioning (\S\ref{sec:adjoint_impl}).
Here, we aim to demonstrate a species of mesh stylization following, e.g., \emph{Cubic Stylization}~\cite{liu2019cubic} or \emph{Homunculus Warping}~\cite{reinert2012homonculus}.
The problem's simplicity makes for a perfect test of the various methods for differentiating through linear solvers.

Stating the problem formally, $\theta$ corresponds to vertex positions $\vec v \in \R^{|\VS| \times 3}$, $\varphi$ is identity, and $\vec t \in \R^{|\VS| \times 3}$ are target vertex positions.
Given a mass matrix as $M$, integrated Laplacian as $L$, and timestep as $\tau$, we define $\vec v' \leftarrow g(\vec v)$ as the result of applying mean curvature fairing to $\vec v$, i.e., solving $(M - \tau L) \vec v' = M \vec v$.
To inflate a mesh, we initialize $\vec t$ and $\vec v$ to the vertex positions of the input mesh; thus, applying mean curvature flow to an optimized $\vec v$ should result in the original vertex positions.
Our implicit relationship $f$ corresponding to this problem is
\begin{equation}
    \vec f(\vec v; \vec v') = (M(\vec v) - \tau L(\vec v)) \vec v' - M(\vec v) \vec v = \vec 0,
\end{equation}
where $L(\vec v)$ and $M(\vec v)$ denote the cotangent Laplacian and diagonal mass matrix, which are computed from the vertex positions $\vec v$. Our \emph{outer} optimization problem is
\begin{equation}\label{eq:invmcf}
    \argmin_{\vec v}\; \frac{1}{2} \lVert \vec t - g(\vec v) \rVert^2_M,
\end{equation}
where $\lVert \vec w \rVert^2_M = \Tr(\vec w^\tpose M \vec w)$ represents the discrete inner product under the $M$-metric for some $\vec w \in \R^{|\VS| \times 3}$.
This problem is (nearly) ill-posed in general, giving some flexibility in how we design the optimization procedure.
Specifically, for a finite number of steps, we can influence the solution by smoothing out the gradients of $\vec v$ using a Sobolev preconditioner as defined in \S\ref{sec:adjoint_impl}.

\begin{figure}[h!]
\centering
\includegraphics[width=\linewidth]{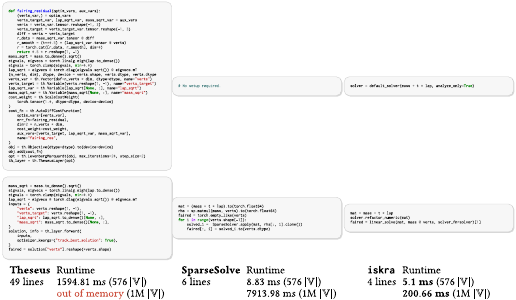}
\caption{Code comparison for inverse fairing. Top row shows the setup needed for each framework, and the bottom compares executing one step of fairing.}\label{fig:fairingcode}
\Description{Figure showing three columns of code. Theseus takes up the most space, \texttt{iskra} the least. \textt{iskra} is also the fastest of the three options. Theseus runs out of memory for a mesh with 1M vertices.}
\end{figure}

This problem demonstrates several features of our system. 
First, because $\vec v$ changes during optimization, we must recompute the Laplacian matrix every iteration.
A naive implementation of the adjoint rule for linear systems would have us refactor the matrix from scratch every step, a potentially costly operation.
However, having domain knowledge of the problem, we could exploit the fact that the sparsity patterns of $M$ and $L$ remain constant throughout by reusing the \emph{symbolic} factorization---i.e., the one that depends on the sparsity pattern alone---if only the system we were using allowed us to do so.
Our system is versatile enough for this purpose, allowing the user to specify a custom linear solver to be used during the forward and backward passes, and manually recompute the \emph{numeric} part of the factorization each iteration, while accounting for $L$'s dependence on $\vec v$.
The \texttt{iskra} code required for this purpose remains compact, as shown in Figure~\ref{fig:fairingcode}.

We offer two points of comparison for this problem.
The first is Theseus~\cite{pineda2022theseus}, a non-linear least squares \emph{declarative} library.
Our inner problem can be easily reformulated as a non-linear least squares and should therefore be solvable by Theseus.
The second is SparseSolve~\cite{jacobson2025sparsesolve}, an imperative library meant exclusively for differentiating through sparse linear systems via the CHOLMOD solver~\cite{Chen:2008:A8C}.
Both natively integrate with PyTorch and support CPU and GPU execution.

\begin{figure}[!ht]
  \centering
  \includegraphics[width=\linewidth]{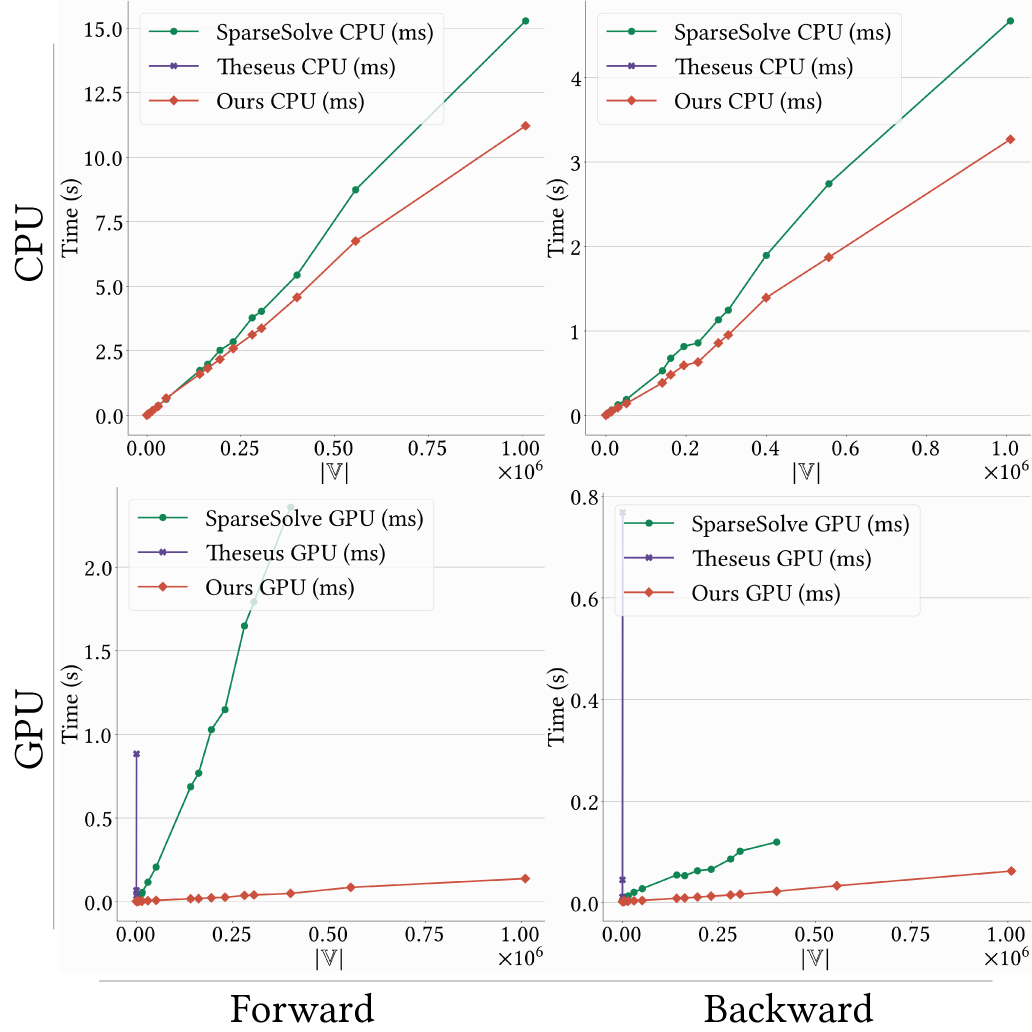}
  \Description{}
  \caption{
  Runtime comparison of linear solver strategies across the CPU and GPU.
  Using a declarative method comes with severe limitation in runtime and memory consumption: Theseus did not succeed at differentiating through systems involving meshes with $>642$ vertices.
  A rigid system that pre-defines a solver is also insufficient for geometry processing as it is often possible to make use of problem-specific strategies.
  }\label{fig:invmcf_time}
\end{figure}

Figure~\ref{fig:fairingcode} visually compares the code required to solve this problem for each of the methods, as well as the execution times for two meshes.
Figure~\ref{fig:invmcf_time} compares the execution speed on a range of meshes as a function of vertex count.
As shown, our system is significantly faster and more ergonomic than Theseus on this problem---implementing it in \texttt{iskra} requires less code and is two orders of magnitude faster on average.
While SparseSolve is comparable to our system in sheer number of lines of code required, and both systems rely on CHOLMOD for CPU execution in this test, \texttt{iskra} is able to exploit problem-specific strategies to further accelerate execution, resulting in an order of magnitude speed up.
Here, \texttt{iskra} is able to seamlessly reuse the factorization in the forward mode and seamlessly switch to the specialized cuDSS solver~\cite{NVIDIA:2025:cudss} for GPU execution.
Our point in comparing performance characteristics of the different solvers is \emph{not} that our system uses a faster solver by default, but that it gives the user flexibility to choose the solution best suited for their problem.

Even though our method is in scope of what Theseus can do, we see that even for such a simple problem, declarative methods can result in a significantly higher implementation effort.
Their system asserts a specific least-squares declarative form for the problem, and, while our problem mathematically fits into this formulation, integrating it with Theseus requires reformulating it: what should amount to a simple sparse linear solve becomes a pipeline involving multiple matrix square roots.

Because they rely on a pre-determined set of solvers, Theseus and SparseSolve cannot exploit specialized solution strategies, such as, e.g., reusing the symbolic part of the factorization. Instead, we are left to rely on one of the pre-selected solvers like Levenberg-Marquardt for Theseus or CHOLMOD with pre-chosen settings for SparseSolve, significantly limiting their performance ceilings. Efficiency issues are far worse for Theseus as it is not designed with sparse geometry operators in mind. Attempting to use a sparse matrix in its problem definition makes the software crash, leaving us to resort to making all operators dense, which in turn negatively impacts memory and speed. In fact, we were unable to use Theseus for meshes larger than $600$ vertices. From this example, we can conclude that due to their rigid implementions of the adjoint rule, previous systems have limited versatility for the problem of inverse fairing. In contrast, our system is designed with flexibility and sparsity in mind, which in turn converts into greater efficiency.

\subsection{Inverse Parameterization}\label{ss:scp}

\begin{figure}[!ht]
  \centering
  \includegraphics[width=\linewidth]{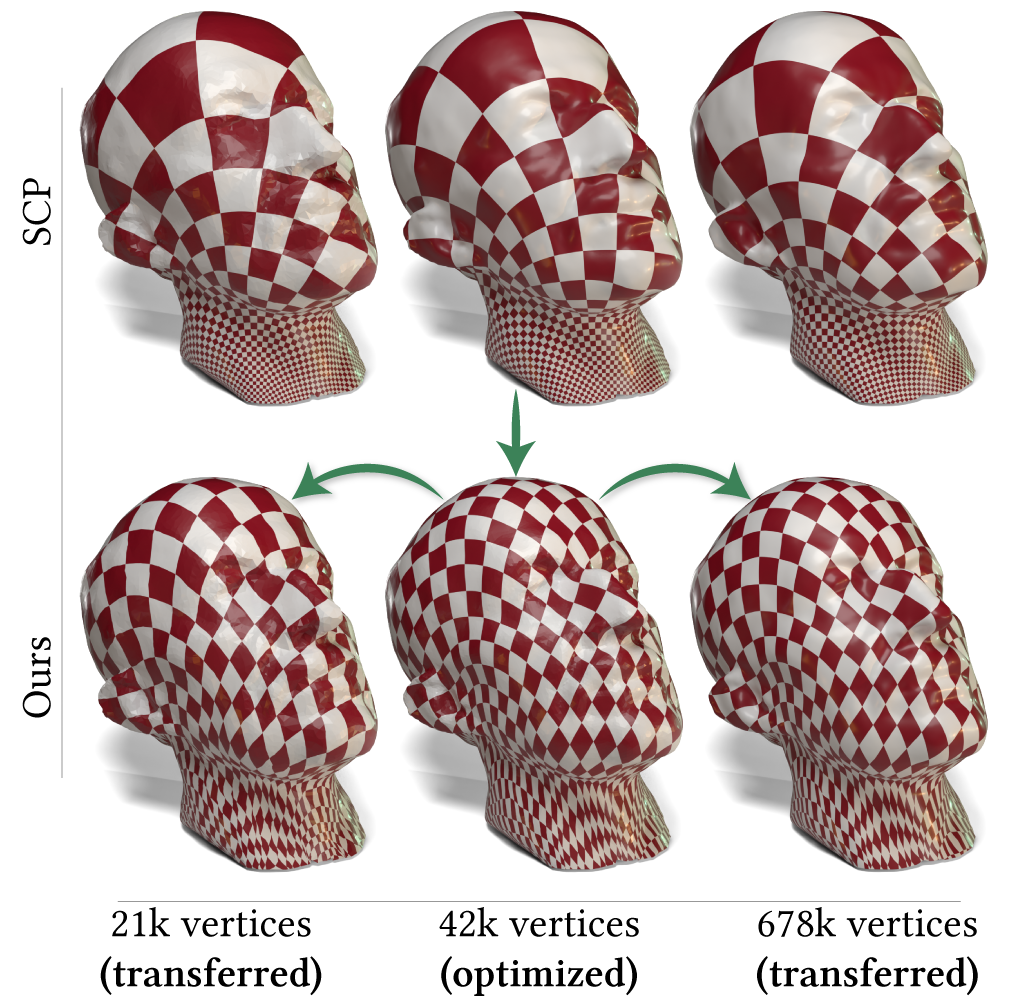}
  \Description{}
  \caption{Left-to-right: same mesh at three different resolutions of a level-of-detail hierarchy. We optimize a neural metric field on the middle-level mesh to reduce area distortion during SCP. Having optimized once, we transfer the metric to other levels by evaluating the field on other meshes.}\label{fig:scp_transfer}
\end{figure}

\begin{figure}[h!]
  \centering
  \includegraphics[width=\linewidth]{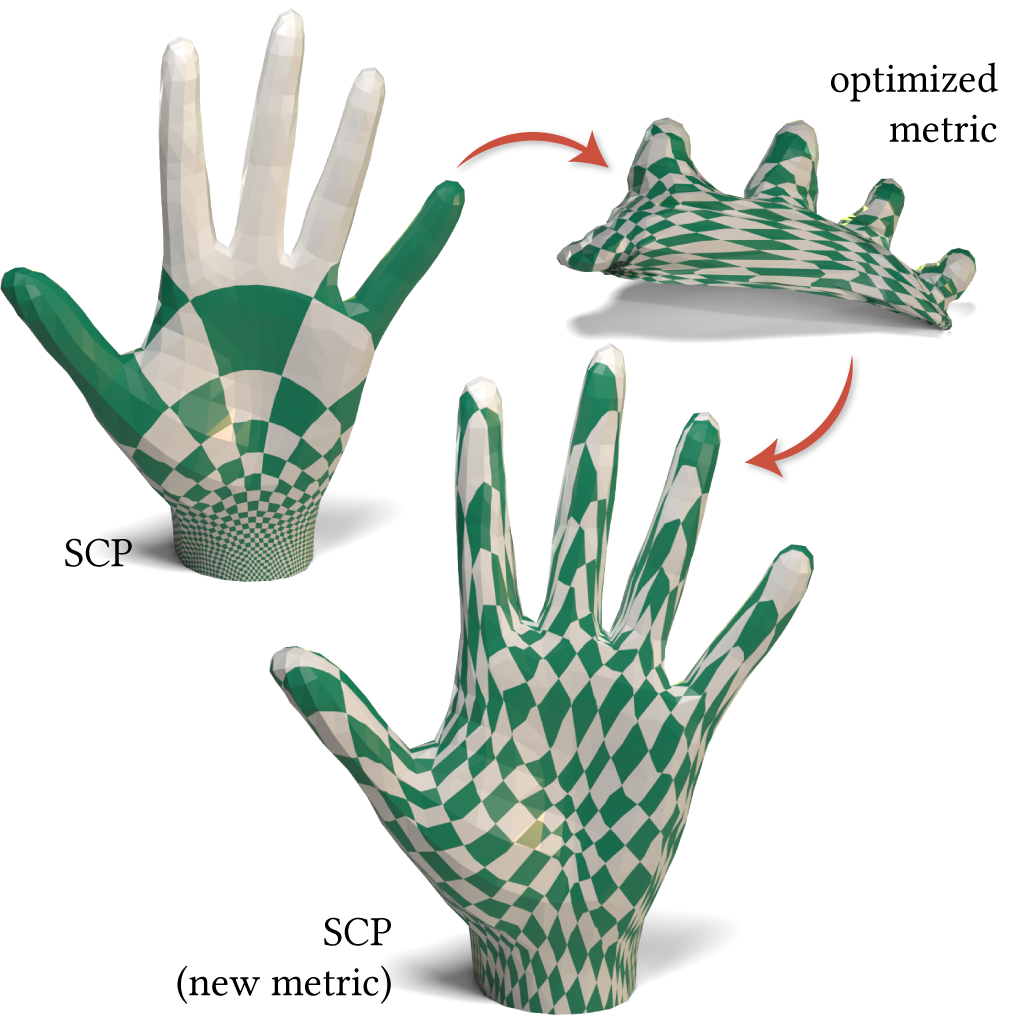}
  \Description{}
  \caption{Spectral conformal parameterizations produce high area distortion. By differentiating through SCP, we can optimize the mesh's metric such that SCP's output under the new metric is less distorted.}\label{fig:scp_explainer}
\end{figure}

Next, we demonstrate our system's capacity for differentiating through sparse eigenproblems.
The spectral conformal parameterization (SCP) algorithm~\cite{Mullen:2008:SCP} has two major benefits over its competitors: it is fast as it only requires solving a sparse eigenproblem, and its solutions are guaranteed to not fold over themselves locally.
However, this comes at the cost of the quality of the map: In Figure~\ref{fig:scp_explainer}, we see that small areas of a texture get assigned to large areas of the mesh, i.e., one pixel on the texture can become stretched over a large area while others are bunched up in a comparatively minuscule region.
Other methods optimize energies meant to reduce this type of distortion, but require additional care to ensure they do not fold over, as well as specialized optimization algorithms~\cite{smith2015bijective,rabinovich2017scalable,wang2023variational}.

Inverse geometry processing offers a way to merge these two approaches.
Our goal is to use SCP as the inner algorithm that maps an input mesh to the plane.
For the outer optimization we minimize some other distortion energy---e.g., the symmetric Dirichlet energy~\cite{schreiner2004inter,smith2015bijective}---by training a neural network to predict an intrinsic metric for SCP to operate in.
In other words, our goal is to design a metric that results in a less distorted parameterization compared to SCP, while retaining SCP's suitable properties.
For our proof-of-concept, we opted for a simple neural field~\cite{xie2022neural} approach, effectively using the neural network as a smooth parameterized interpolant and optimizing on one specific mesh.

This application illustrates how our system enables new pipelines driven by ambient neural fields.
Unlike traditional methods, using a neural field allows us to transfer the optimized metric onto a remeshed version of the same geometry, e.g., transferring the parameterization down a level-of-detail mesh hierarchy as shown in Figure~\ref{fig:scp_transfer}, a setting not handled by existing approaches.
While the purpose of our experiment is this new setting (rather than outperforming the single-mesh parameterization literature), we note that some parameterization methods already do solve \emph{linear} inverse problems, e.g., \citet{wang2023maps}, without being transferrable.

SCP is an intrinsic algorithm, meaning it takes as input a set of edge lengths $\vec l \in \R^{|\ES|}$ that satisfy certain properties, such comprising a discrete metric with no degenerate elements, and outputs a set of 2D coordinates $\vec u \in \R^{|\VS| \times 2}$; thus, we write $\vec u \leftarrow g(\vec l)$.
In our example, $\theta$ are the parameters of a neural field operating in ambient space that we will use to predict a new discrete metric.
A simple approach is to predict a displacement vector per mesh vertex and use the displaced vertex positions to compute the discrete metric.
Therefore, our neural network takes as input a point in ambient space $\vec p \in \R^3$ and outputs an offset $\vec o \in \R^3$: for edge $i$ connecting vertices $k$ and $l$, the new edge lengths are $l'_{i} = \lVert (\vec p_j + \vec o_j) - (\vec p_k + \vec o_k)\rVert$.
Since we only ever evaluate our neural network on the mesh vertex positions, we can view the composition of its evaluation with the computation of the edge lengths as $\varphi : \theta \mapsto l'$.
We use a multilayer perceptron with $3$ hidden layers of $64$ neurons each, and use a harmonic input encoding~\cite{mildenhall2020nerf} with $5$ basis functions.

Given $\vec l'$, SCP constructs a discrete \emph{conformal} Laplacian $L_c \in \R^{2|\VS| \times 2|\VS|}$ and solves the generalized eigenproblem
\begin{equation}
    L_c \, \vectorize(\vec u) = \lambda\, B\, \vectorize(\vec u),
\end{equation}
where $B$ is SCP's diagonal ``boundary matrix,'' and $\vectorize(\cdot)$ flattens $\vec u$ into a $\R^{2|\ES|}$ vector.
Our parameterization is then the eigenvector $\vec u$ corresponding to the smallest non-zero eigenvector; for more information, refer to~\citet{Mullen:2008:SCP}.
In code, the user computes the cotangent weights and calls \texttt{iskra}'s spectral implicit layer:
\begin{minted}{python}
def scp(lengths, face_edge, n_vertices, faces, bdr_idx, boundary_mat):
    cot = cotan_weights_intrinsic(lengths, face_edge)
    conf_lap = conformal_laplacian(cot, n_vertices, faces, bdr_idx)
    _, evecs = eigsh(
        conf_lap,
        M=boundary_mat,
        k=3,
        sigma=-1e-12,
        bwd_method="fixed-point",
        bwd_max_iter=25,
    )
    uv = evecs[..., 0:1].reshape(2, -1).mT
    return uv
\end{minted}
\texttt{iskra} implements the forward pass via a sparse generalized eigensolver and allows the user to choose one of a number of different approaches for computing the backward pass; see below for a comparison of different methods.
In practice, we precompute and cache purely topological matrices needed to construct the conformal Laplacian, excluding them here for the sake of exposition.

Our outer optimization objective is the well known symmetric Dirichlet energy~\cite{schreiner2004inter,smith2015bijective}.
Given the Jacobian $J_i$ of the map between the rest configuration of triangle $i \in \FS$ and its parameterization, its rest area $\vec a_i$, we can write the symmetric Dirichlet energy as:
\begin{equation}
    \ell(\vec l') = \sum_{i \in \FS} \vec a_i (\lVert J_i \rVert_{F} + \lVert J_j^{-1} \rVert_{F}),
\end{equation}
This can be implemented in \texttt{iskra} via a gather operation followed by standard PyTorch linear algebra.
We optimize this objective using the Adam optimizer with an initial learning rate of $10^{-3}$, for a total of $2000$ steps.
Unlike other methods for optimizing the symmetric Dirichlet energy~\cite{smith2015bijective,rabinovich2017scalable}, our method requires no specialized line-search procedure during gradient descent as going through SCP guarantees injectivity.

\begin{figure}[h!]
  \centering
  \includegraphics[width=\linewidth]{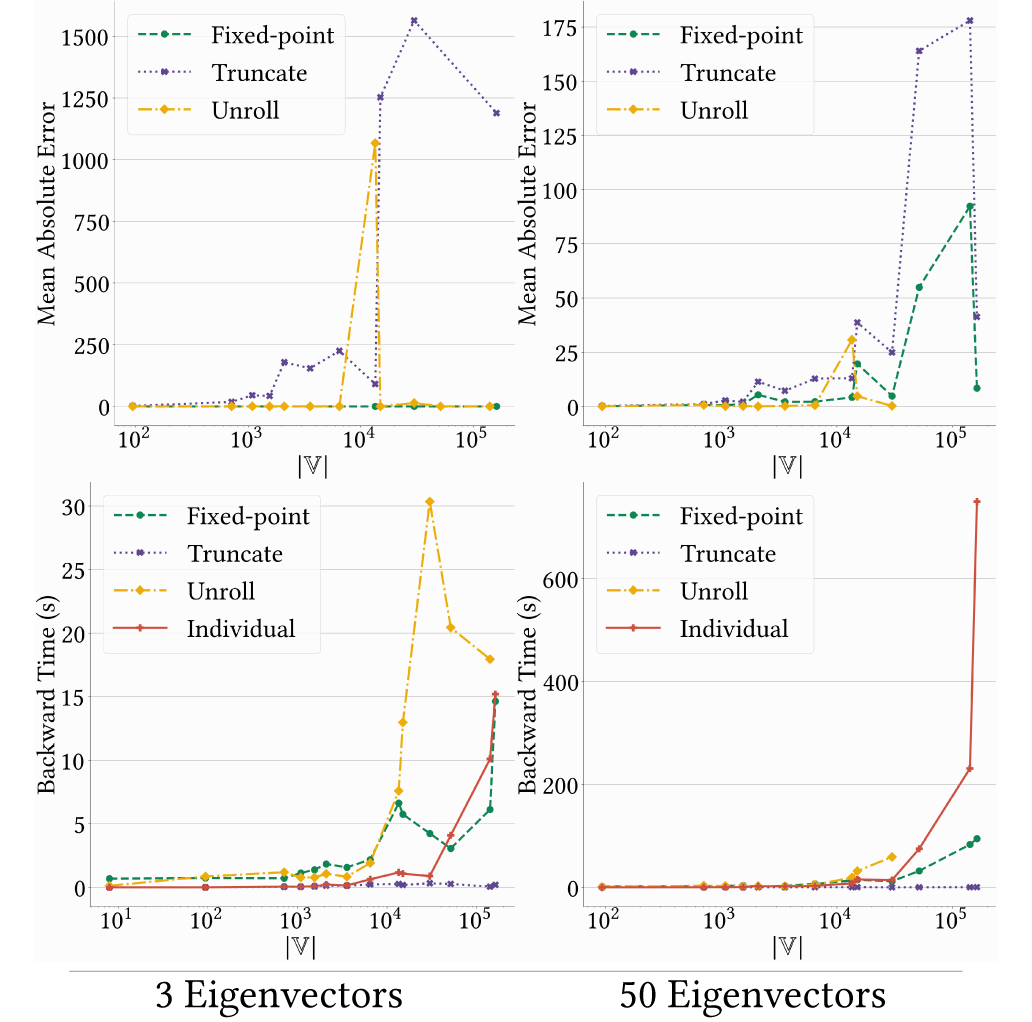}
  \Description{}
  \caption{Comparison of eigenvector derivative computation methods across mesh scales and number of eigenvectors. Top row compares accuracy with respect to \emph{Individual}. Missing data on for the \emph{Unroll} method is due to the fact that stepping back through power iterations requires more memory than available on our system for large enough problems.}\label{fig:eigenstudy}
\end{figure}

\paragraph*{Eigenstudy}
Our system implements four methods for differentiating through eigensystems.
We compare the methods in Figure~\ref{fig:eigenstudy} in terms of runtime and accuracy of differentiating through Laplacian eigenvectors, as functions of mesh size and the number of requested eigenvectors.

The first algorithm directly implements Equation~\ref{eq:eigindividual}, solving a different linear system for each eigenvector requested by the user.
We treat this method as ground truth for accuracy as it involves no approximations.
While it is fast for a small number of eigenvalues, the runtime explodes for a large number of eigenvectors and/or a large mesh, especially since this method has to refactor a large matrix for each requested eigenvector.

The simplest approach \emph{unrolls} power iterations, i.e., stepping back through each iteration of the power iteration algorithm.  For our experiments, we used $100$ iterations. 
While fairly accurate, this method requires high memory use; we were unable to run it on meshes with more than $30$K vertices for three eigenvectors, and on meshes larger than $162$K vertices for $50$ eigenvectors.
This makes this method of limited practical usefulness, especially since the isolated workload in this test case was not integrated into a larger computational pipeline.

Next is the method of \citet{smirnov2021hodgenet}, which \emph{truncates} the number of eigenvalues used to evaluate the closed-form Equation~\ref{eq:eigensmirnov}.
The method is the fastest as it only inverts one sparse matrix.
However, this comes at the cost of accuracy if the number of requested eigenvectors is small relative to the mesh size; conversely, it is well suited for problems where the user requests many eigenvectors relative to mesh size.

Lastly, we offer an algorithm that relies on our \emph{fixed-point} functionality to differentiate through the QR relationship in Equation~\ref{eq:eigiter}.
We set the number of $GMRES$ iterations to $100$and GMRES tolerances to $0$ to match the number of unrolled power-iterations in the unrolled setting.
Figure~\ref{fig:eigenstudy} shows that this method offers a good middle ground: it is comparably accurate to other approximate methods and is strictly more accurate than truncation on all but one mesh.
While it is slower than truncation, it is comparable to the remaining methods on small meshes and faster on large meshes across both test conditions.
In comparison to the \emph{individual} method, \emph{fixed-point} can pre-factor once and reuse the solver across iterations.

As evidenced by our experiments, unrolling power iterations is impractical, \emph{individual} differentiation should be used for small meshes and small numbers of eigenvectors, \emph{truncated} differentiation should be used for large numbers of eigenvectors, and \emph{fixed-point} differentiation offers a reasonable trade-off.
All experiments were performed on the CPU.

\subsection{Inverse ARAP}\label{ss:arap}

\begin{figure}[h]
  \centering
  \includegraphics[width=\linewidth]{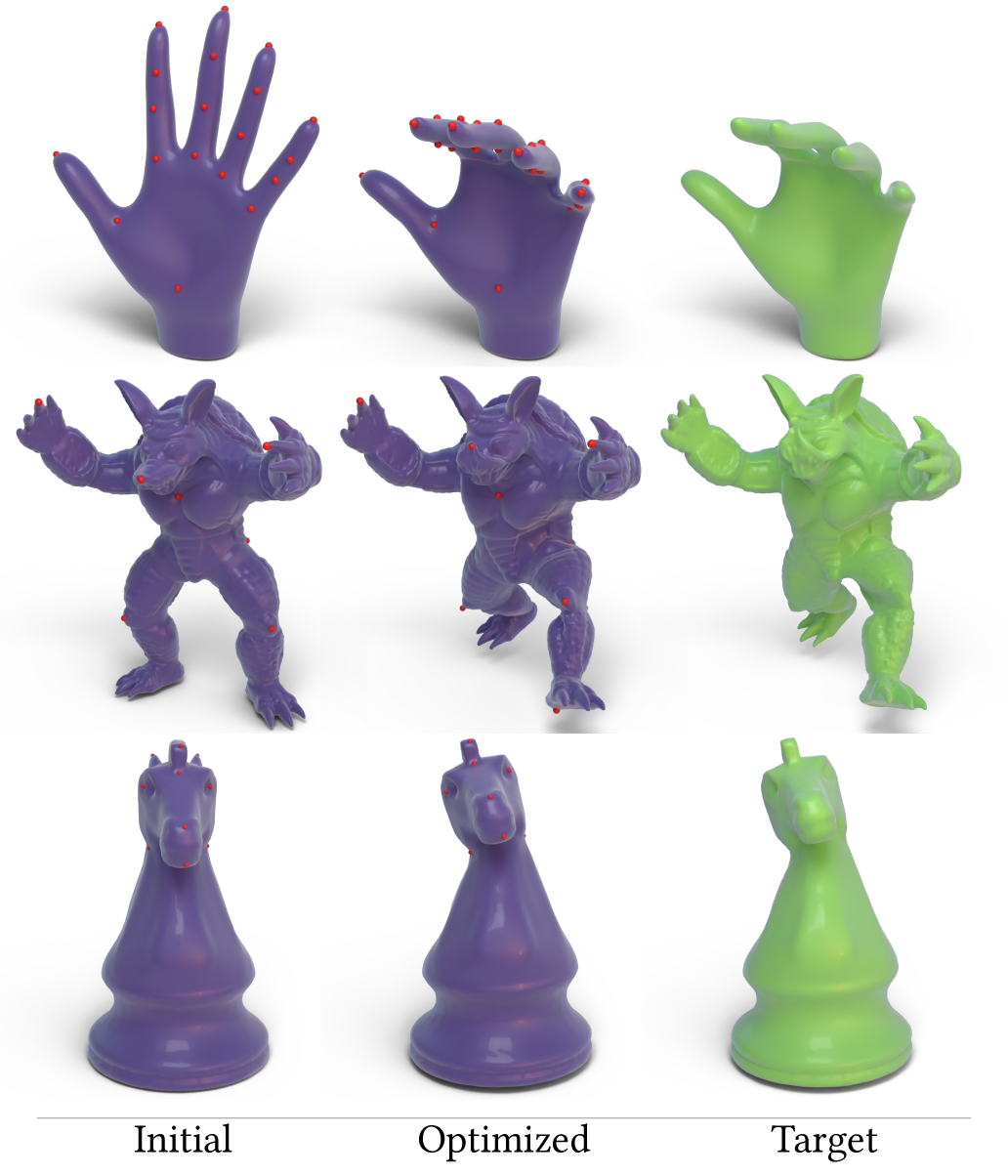}
  \Description{}
  \caption{Results of using our system to solve the inverse ARAP problem on a range of meshes.}\label{fig:arap}
\end{figure}

We now return to our introductory example of inverting the ARAP problem.
Formally, assuming we have cotangent Laplacian weights $\vec c$, the optimization problem solved by ARAP is defined as:
\begin{equation}
\begin{aligned}
\argmin_{\vec v' \in \R^{|\VS| \times 3}, \{R_1, \ldots, R_{|\VS|}\}} \quad & \sum_{(i, j) \in \VV} \vec c_{ij} \left\| \left(\vec v'_i - \vec v'_j\right) - R_i(\vec v_i - \vec v_j) \right\|^2 \\
\suchThat \quad & \vec v'_k = \vec h_k, \; k \in \mathbb H \\
& R_i \in SO(3), \; i = 1, \ldots, |\VS|,
\end{aligned}
\end{equation}
where $\mathbb H$ is the set of handles with positions $\vec h$ and $R_i$ are per-vertex rotation matrices.
We have already shown how to implement the specialized local-global ARAP solver of \citet{Sorkine:ARAP} using our system's fixed-point interface and use it to solve the inverse problem from Equation~\ref{eq:invarapprob} (\S\ref{ss:defimplicit}).
We demonstrate the results of this optimization on number of meshes in Figures~\ref{fig:teaser}~and~\ref{fig:arap}.
All results were computed by running the outer optimization loop for $500$ iterations.
The number of GMRES iterations was kept at 200.

\begin{figure}[t]
  \centering
  \includegraphics[width=\linewidth]{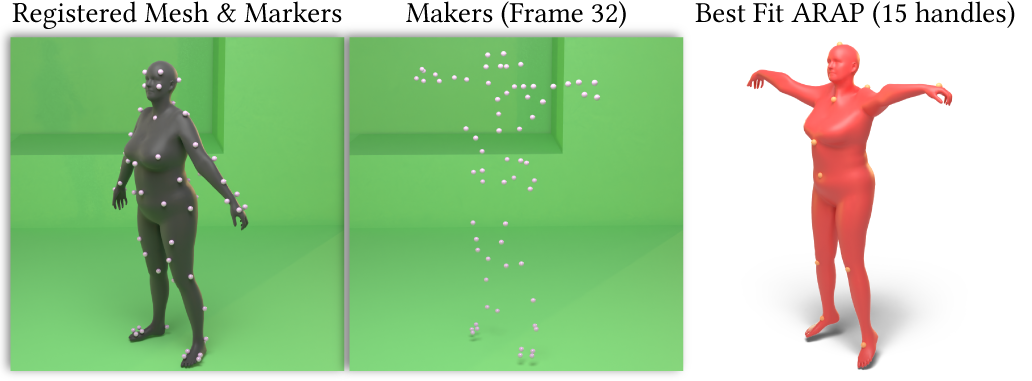}
  \Description{}
  \caption{Fitting an ARAP deformation to match motion capture markers.}\label{fig:amass}
\end{figure}

We also include a modification to the problem in Figure~\ref{fig:amass}, which uses motion capture data from the AMASS dataset~\cite{AMASS}.
First, we take motion capture markers on the registered body mesh in rest pose and find the input mesh vertex closest to each marker.
The goal is to match those vertices to markers captured during the person jumping by fitting a reduced order deformation model to the new set of markers.
We compute the loss in \eqref{eq:invarapprob} between this subset of vertices and the markers, instead of on the full mesh.
Notably, the number of ARAP handles is much smaller than the number of markers.

ARAP can be relaxed into a nonlinear least squares problem~\cite{mara2021thallo}, allowing us to compare with Theseus~\cite{pineda2022theseus} once again.
Unlike in \S\ref{ss:mcf}, ARAP's nonlinear least squares formulation can naturally be written without sparse matrices, meaning we manually convert our geometry operators into dense operations.
Table~\ref{tbl:arap_theseus} includes a comparison on a mesh of a sphere with $642$ vertices: we ran both implementations for a total of $25$ optimization steps and report the forward and backward pass runtimes, as well as the peak memory use across both CPU and GPU back-ends.
\begin{table}[h!]
\caption{Despite ARAP falling under the class of problems solvable by Theseus, \texttt{iskra}'s imperative approach remains orders of magnitude faster. Note Theseus did not complete executing on larger meshes due to excessive memory requirements.}\label{tbl:arap_theseus}
\begin{NiceTabularX}{\linewidth}{>{\normalsize}l>{\normalsize}X>{\normalsize}r>{\normalsize}r>{\normalsize}r}
  \CodeBefore%
  \rowcolors{2}{gray!10}{}[restart,respect-blocks]
  \Body%
  \toprule
  & & Forward (s) & Backward (s) & Memory (GB)\\
  \midrule
  \!\Block{2-1}{CPU}\!& Theseus & $43.69$ & $135.38$ & $18.22$ \\
  & \texttt{iskra} & $0.32$ & $0.027$ & $0.67$ \\
  \midrule
  \!\Block{2-1}{GPU}\!& Theseus & $1.18$ & $2.39$ & $13.76$ \\
  & \texttt{iskra} & $0.095$ & $0.018$ & $0.019$ \\
  \bottomrule
\end{NiceTabularX}
\end{table}
Here we see how our system's versatility leads to greater efficiency.
As expected, being able to use the specialized algorithm of \citet{Sorkine:ARAP} as opposed to a generic solver accelerates the forward pass.
Moreover, we see how the generality of our system leads to orders of magnitude better memory characteristics and an order of magnitude faster backward pass. 
We were unable to run Theseus on meshes significantly larger than $642$ vertices as its memory allocations exceeded our computer's $32$GB or RAM memory.

\subsection{Inverse Geodesic Distances}\label{ss:geodesics}
\begin{figure}[t]
  \centering
  \includegraphics[width=\linewidth]{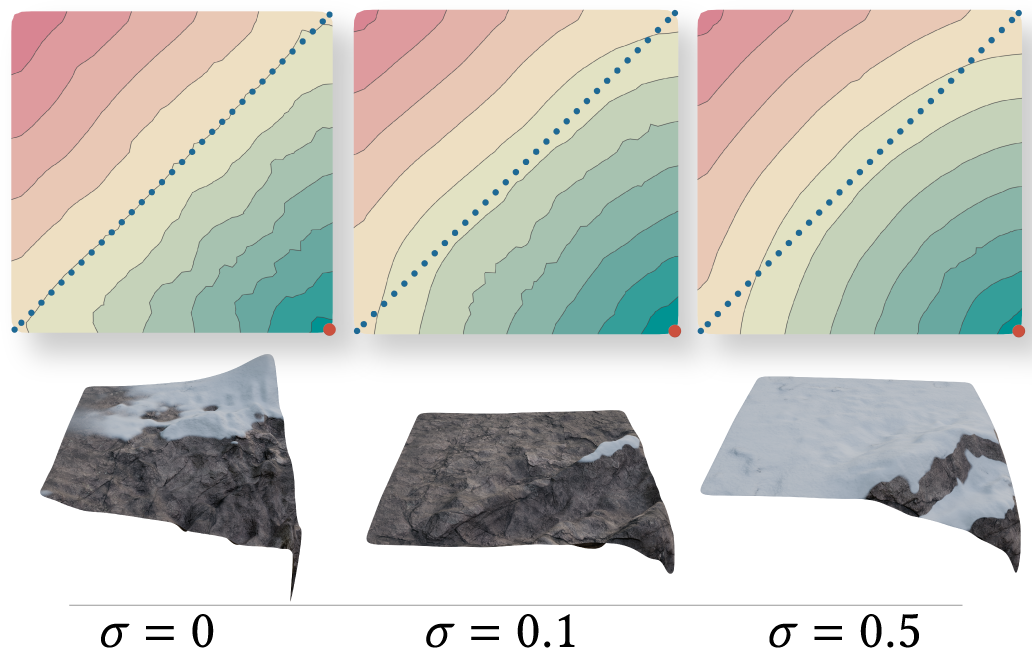}
  \Description{}
  \caption{Our system can differentiate through regularized geodesic distances, facilitating procedural terrain design. Plots in the top row show distances to their bottom-right corners (red). These correspond to a birds-eye view of the heightfield in the bottom row. The objective is to make the centerline points (blue) equidistant from the start point, in combination with a smoothness regularizer governed by $\sigma$. Different columns correspond to different regularization strengths.}\label{fig:terrain}
\end{figure}

In this section, we use our system's fixed-point implicit layers to differentiate through geodesic distances, in the process demonstrating our system's support for the large class of geometry processing problems solvable via ADMM iterations~\cite{boyd2011distributed,glowinski1975approximation,gabay1976dual}
While prior work relied on hand-derived formulae for differentiating through geodesic distances~\cite{benmansour2010geodesics,li2024geodesics}, our method automates this derivation for its users.
In the \emph{inner} problem, we are optimizing for a \emph{regularized} distance function $u \in \R^{|\VS|}$ starting from some set of vertices $\mathbb S \subset \VS$.
It is regularized because it trades off satisfying the geodesic equations and, in our case, the Dirichlet energy via a chosen parameter $\beta$,
Assuming vertex areas $m \in \R^{|\VS|}$, and the discrete gradient operator $G$ and Laplacian $L$, we can phrase this problem (with an optional smoothing term controlled by $\beta\geq0$) as follows
\begin{equation}\label{eq:edelstein}
\begin{aligned}
\min_u \quad & -m^\tpose u + \frac{\beta}{2} u^\tpose L u \\
\suchThat \quad & \lVert (Gu)_f \rVert \leq 1, \quad \forall f \in \FS \\
& u_s = 0, \quad \forall s \in \mathbb S.
\end{aligned}
\end{equation}
As with previous examples, geometry processing practitioners have proposed specialized solvers that exploit problem-specific structure---in this case a fast and concise ADMM scheme proposed by~\citet{edelstein2023rgd}, who introduced formulation~\eqref{eq:edelstein}.
This scheme alternates three steps: solving for $u$, solving for a auxiliary variable $z = Gu$, and solving for the dual variable $\lambda \in \R^{|\FS|}$.
Assuming a set of algorithm specific constants $\rho$ and $\beta_k$, and the divergence operator $D$, this procedure can be compactly implemented in in \texttt{iskra} as:
\begin{minted}{python}
# step 1: u-minimization
b = vert_areas - div @ y.flatten() + rho * div @ z.flatten()
u = linear_solve(lap, b, solver=solver)[1] / (beta + rho)

# step 2: z-minimization
grad_u = (grad @ u).reshape(*z.shape)
z_new = (1 / rho) * y + grad_u
norm_z = torch.linalg.vector_norm(z_new, axis=0)
z_new = z_new / torch.where(norm_z >= 1, norm_z, 1)

# step 3: dual update
y_new = y + rho * (beta_k * grad_u + (1 - beta_k) * z - z_new)
\end{minted}
In practice, we also implemented the computation of primal and dual residuals, heuristic $\rho$ modulation, and early termination criteria proposed in~\cite{edelstein2023rgd}, bringing up the number of lines of code from $7$ to $25$ for the inner loop.
The user defines $g$ by passing this code to our fixed-point interface (\S\ref{ss:defimplicit}), specifying which variables act as iterates and which outputs represent the residuals.%

For our \emph{outer} optimization, we procedurally design a terrain, e.g., for a videogame.
Our mesh is a flat plane embedded in 2D, and our optimization variables are the heights of a heightfield defined on that mesh, $\theta \coloneqq \zeta \in \R^{|\VS|}$; $\varphi$ is the identity map.
We design the terrain such that the distance a character needs to traverse from its corner to any point on its diagonal equals some constant $u_c$.
Our objective is composed from $3$ terms: a distance term, a smoothness term weighed by a factor $\sigma$, and a non-negativity term:
\begin{equation*}
\begin{aligned}
    \argmin_{\zeta}\; \underbrace{\sum_{i \in \mathbb C} (u_c - g(\zeta)_i)^2}_{\text{center-distance}} + \underbrace{\sigma \zeta^\tpose L \zeta}_{\text{smoothness}} + \underbrace{0.1 \sum_{i \in \VS} \relu(-\log(\zeta_i + 1))}_{\text{non-negativity}},
\end{aligned}
\end{equation*}
where $\mathbb C$ denotes the set of vertices on the heightfield's diagonal.
We optimize the outer loop using $H^1$ descent while constraining $\zeta_{s} = 0$ for all $s \in \mathbb S$ to eliminate translational ambiguity.
We show the results produced by our system after $500$ iterations in Figure~\ref{fig:terrain} for a range of smoothness values $\sigma$.

This problem is a quadratic program, solvable systems designed for solving and differentiating through this broad class of problems. %
Here we include a comparison with the seminal CVXPYLayers system~\cite{agrawal2019cvxpylayers} which builds on the CVX optimization package~\cite{cvx}, running all methods with their default parameters.
Implementing this problem in CVXPYLayers required us to reformulate the problem in two ways.
First, it required us to manually rewrite sparse matrix operations in terms of dense tensor operations.
Second, we had to convert our problem into the so-called disciplined parameterized programming paradigm of CVXPYLayers~\cite{agrawal2019cvxpylayers}.
This is a non-trivial exercise, requiring the user to reason about the ``affineness'' of each operation in the inner loop---for more detail see \cite{agrawal2019cvxpylayers}.

CVXPYLayers successfully solves this problem on a terrain mesh with $2048$ faces.
Both methods acheived a comparable loss on the outer problem after $500$ iterations with a mutual difference of $10^{-4}$.
Due to the problem reformulation necessitated by CVXPYLayers, the amount of code required by both methods ended up being comparable, and so were the forward pass runtimes. 
On the CPU our forward pass on average took $0.299$s (ranging from $0.150$s to $1.163$s), compared to CVXPYLayers $0.301$s ($0.194$s to $0.661$s).
However, our backward pass was faster: on average our method took $3.670$s ($0.150$ to $6.919$s), whereas CVXPYLayers took $12.335$s, ($5.538$ to $13.407$s).
The variation in the runtimes are due to the early stopping critaria of both methods, and, in our case, of GMRES during the backward pass.
While comparable on sheer runtime, our method outperforms CVXPYLayers by an order of magnitude when it comes to memory use.
Simply applying one step of Loop subdivision~\cite{Loop1987SmoothSS} results in CVXPYLayers attempting to allocate more memory than available on our $32$GB machine, leading it to crash.
In contrast, our system's memory usage remains below $1$GB for this test case.

\section{Discussion}\label{s:discussion}

Our results thoroughly demonstrate our system in relation to its design goals of versatility, usability, interoperability, and efficiency.
In particular, we measure versatility by solving problems involving sparse linear systems, sparse eigendecompositions, local-global iterations, and ADMM formulations, each representative of a large class of geometry problems.
As shown in \S\ref{ss:mcf} and \S\ref{ss:arap}, given a formulation of the desired implicit relationship $f$, \texttt{iskra}'s code is often more succinct than its declarative counterparts, demonstrating its usability.
Experiments like the one in \S\ref{ss:scp} showcase interoperability by directly integrating with machine learning architectures through PyTorch.
In all of our experiments, \texttt{iskra} is either comparable to, or strictly faster than alternatives.

Design goals like ours are often in tension with one another.  For example, usability typically inhibits low-level control, which otherwise would imply greater implementation effort.  Surprisingly, however, our system satisfies all the design goals simultaneously.  
In fact, succinctly expressing complex geometric algorithms \emph{necessitated} additional affordances for simplicity and expressivity.
These affordances include our tensor-based hierarchical scatter-gather formulation, sparse linear algebra functionality, and high-level utilities for fixed-point iteration algorithms. 
These features are crucial for geometry processing, allowing practitioners to differentiate through existing specialized algorithms without needing to reformulate them.

Moreover, we have demonstrated how our system naturally integrates with machine learning workflows (\S\ref{ss:scp}), opening doors to a broad class of geometry processing pipelines.
While our example in \S\ref{ss:scp} offers a proof-of-concept integration with a neural field, future directions may combine implicit geometric layers with larger-scale data-driven machine learning workflows.
Extrapolating from this example, a future data-driven model may provide priors and then rely on implicit geometric layers to guarantee that geometric constraints are satisfied.
This direction is particularly compelling as it promises to marry two, until now, mostly disjoint worlds in geometry processing: data-driven and optimization-based methods.

Despite our framework's simplicity and concision in practical examples, we achieve versatility by requiring the user to provide an implicit relationship.
In practice, this is desirable because we wish to use existing geometry processing formulations.
However, one could imagine adding utilities similar to our fixed-point interface, e.g., one that translates a declarative-style constrained quadratic program specification into an implicit function (like CVX~\cite{cvx,gb08}) or another that creates a proximal iteration to be plugged into our fixed-point interface (like $\nabla$-Prox~\cite{lai2023deltaprox}).
While we focused on existing geometry processing algorithms, these extensions may enable rapid prototyping of completely new geometry processing techniques.

Furthermore, %
one could likely extend our system to physics simulation, 
taking inspiration from, e.g., \citet{huang2024diff}.
We suspect that implementing collision detection or friction could be accomplished using differentiable utilities in the geometry module, orthogonal to the rest of our system---e.g., \S\ref{ss:geodesics} already implements a na\"ive penalty forcing terrain heights to be nonnegative.
Differentiating through ordinary differential equations is currently possible either by unrolling multiple implicit layers or by specifying the system within our fixed-point interface; we leave for future work implementing time integrators within our custom backward pass interface.
Lastly, we only handle dynamic topology changes \emph{outside} the implicit function interface: the vector-Jacobian product used to construct the backward pass does not see topology changes that may have happened during the forward pass. 
Differentiable topology changes (see, e.g., \citet{rakotosaona2021difftri}) remain an avenue for future work.

While the generality of the off-the-shelf GMRES optimizer allows us to differentiate through a broad class of problems, future work could look into faster or more robust approaches.
For systems with multiple solutions, GMRES produces the ``Drazin inverse'' solution, making it somewhat more predictable than na\"ive direct solvers.
Singular Jacobians are still possible, but we did not encounter any empirically.
We did experience poorly-conditioned Jacobians, e.g., in the geodesics example, where many GMRES iterations were necessary for convergence.
In this case, we encountered the warning stating GMRES failed to converge within the chosen iteration count---since gradient descent can recover from minor gradient perturbations, however, we did not ultimately find this to be an issue.
While \texttt{iskra} allows for custom solvers for the backward pass,  Jacobi preconditioning was overall slower despite needing fewer iterations; we leave derivation of more effective preconditioning strategies to future work.

While not explored here, our method can also be combined with other differentiation utilities in geometry processing, such as libraries that allow users to construct sparse Hessians from energies defined on meshes~\cite{jacobson2026indexed}.
Our system may also benefit from future innovation in differentiable mesh systems that bridge the gap between low-level high-performance GPU data structures~\cite{yu2023taichi,mahmoud2021rxmesh} and tensor-based machine learning systems.
Moreover, we foresee possible extensions and integrations with differentiable physics simulation and contact handling~\cite{warp2022,hu2019difftaichi}, higher-order or non-simplicial finite elements~\cite{polyfem}, and algorithms that involve dynamic topology changes~\cite{huang2024diff}.
Lastly, our system is likely compatible with inverse rendering~\cite{NimierDavidVicini2019Mitsuba2,pytorch}, enabling end-to-end differentiable graphics applications.

In providing a system for the automatic computation of adjoints, we bring the same kind of simplicity and ease of experimentation to inverse problems in geometry processing that automatic differentiation frameworks have brought to other domains.
Our system opens doors to new geometry processing pipelines, unifying machine learning approaches with classical optimization-based algorithms.
It does so in a compact, fast, memory efficient manner, without requiring practitioners to reformulate their algorithms.

\begin{acks}
We thank Brandon Amos, Oded Stein, and Jonathan Ragan-Kelley for helpful early discussions.
The MIT Geometric Data Processing Group acknowledges the generous support of National Science Foundation grants IIS2335492 and OAC2403239, from the CSAIL Future of Data and FinTechAI programs, from the MIT--IBM Watson AI Laboratory, from the Wistron Corporation, from the MIT Generative AI Impact Consortium, from the Toyota--CSAIL Joint Research Center, and from Schmidt Sciences.
\end{acks}

\bibliographystyle{ACM-Reference-Format}
\bibliography{bibliography}

\end{document}